\documentclass[preprint]{aastex}
\usepackage{graphicx}

\begin{document}

\title{CALIBRATIONS OF ATMOSPHERIC PARAMETERS OBTAINED FROM THE FIRST YEAR OF SDSS-III APOGEE OBSERVATIONS }

\author{
Sz.~M{\'e}sz{\'a}ros\altaffilmark{1,2}, 
J.~Holtzman\altaffilmark{3},
A.~E.~Garc\'{\i}a P\'erez\altaffilmark{4},
C.~Allende~Prieto\altaffilmark{1,2},
R.~P.~Schiavon\altaffilmark{5}, 
S.~Basu\altaffilmark{6}, 
D.~Bizyaev\altaffilmark{7}, 
W.~J.~Chaplin\altaffilmark{8}, 
S.~D.~Chojnowski\altaffilmark{4},
K.~Cunha\altaffilmark{9,10},
Y.~Elsworth\altaffilmark{8}, 
C.~Epstein\altaffilmark{11}, 
P.~M.~Frinchaboy\altaffilmark{12},
R.~A.~Garc\'{\i}a\altaffilmark{13},
F.~R.~Hearty\altaffilmark{4},
S.~Hekker\altaffilmark{14}, 
J.~A.~Johnson\altaffilmark{11}, 
T.~Kallinger\altaffilmark{15},
L.~Koesterke\altaffilmark{16},
S.~R.~Majewski\altaffilmark{4},
S.~L.~Martell\altaffilmark{17}, 
D.~Nidever\altaffilmark{18},
M.~H.~Pinsonneault\altaffilmark{11},
J.~O'Connell\altaffilmark{12},
M.~Shetrone\altaffilmark{19},
V.~V.~Smith\altaffilmark{20},
J.~C.~Wilson\altaffilmark{4},
G.~Zasowski\altaffilmark{11}
}

\altaffiltext{1}{Instituto de Astrof{\'{\i}}sica de Canarias (IAC), E-38200 La Laguna, Tenerife, Spain}
\altaffiltext{2}{Departamento de Astrof{\'{\i}}sica, Universidad de La Laguna (ULL), E-38206 la Laguna, Tenerife, 
Spain}
\altaffiltext{3}{New Mexico State University, Las Cruces, NM 88003, USA}
\altaffiltext{4}{Department of Astronomy, University of Virginia, Charlottesville, VA 22904-4325, USA}
\altaffiltext{5}{Astrophysics Research Institute, Liverpool John Moores University, Egerton Wharf, Birkenhead, 
Wirral. CH41 1LD, United Kingdom}
\altaffiltext{6}{Department of Astronomy, Yale University, P.O. Box 208101, New Haven CT 06520-8101, USA}
\altaffiltext{7}{Apache Point Observatory, PO Box 59, Sunspot, NM 88349-0059, USA}
\altaffiltext{8}{University of Birmingham, School of Physics and Astronomy, Edgbaston, Birmingham B15 2TT, United
Kingdom}
\altaffiltext{9}{Steward Observatory, University of Arizona, Tucson, AZ 85721, USA} 
\altaffiltext{10}{Observat\'orio Nacional, S\~ao Crist\'ov\~ao, Rio de Janeiro, Brazil}
\altaffiltext{11}{Department of Astronomy, The Ohio State University, Columbus, OH 43210, USA}
\altaffiltext{12}{Texas Christian University, Fort Worth, TX 76129, USA}
\altaffiltext{13}{Laboratoire AIM, CEA/DSM-CNRS-Universit\'e Paris Diderot; IRFU/SAp, Centre de Saclay, 
91191 Gif-sur-Yvette Cedex, France}
\altaffiltext{14}{Astronomical institute `Anton Pannekoek', University of Amsterdam, Science Park 904, 
1098 XH, Amsterdam, the Netherlands}
\altaffiltext{15}{Institute for Astronomy, University of Vienna, T\"urkenschanzstrasse 17, A-1180 Vienna, Austria}
\altaffiltext{16}{Texas Advanced Computing Center, University of Texas, Austin, TX, 78759}
\altaffiltext{17}{Australian Astronomical Observatory, 105 Delhi Road, North Ryde NSW 2113, Australia}
\altaffiltext{18}{Department of Astronomy, University of Michigan, Ann Arbor, MI 48109, USA} 
\altaffiltext{19}{University of Texas at Austin, McDonald Observatory, Fort Davis, TX 79734, USA}
\altaffiltext{20}{National Optical Observatories, Tucson, AZ 85719, USA}

\begin{abstract}

The SDSS-III Apache Point Observatory Galactic Evolution Experiment
(APOGEE) is a three year survey that is collecting 10$^5$
high-resolution spectra in the near-IR across multiple Galactic
populations. To derive stellar parameters and chemical
compositions from this massive data set, the APOGEE Stellar Parameters
and Chemical Abundances Pipeline (ASPCAP) has been developed.
Here, we describe empirical calibrations of stellar parameters
presented in the first SDSS-III APOGEE data release (DR10).  These
calibrations were enabled by observations of 559 stars in 
20 globular and open clusters.  The
cluster observations were supplemented by observations of stars in
NASA's \textit{Kepler} field that have well determined surface gravities
from asteroseismic analysis. We discuss the accuracy and precision
of the derived stellar parameters, considering especially effective
temperature, surface gravity, and metallicity; we also briefly
discuss the derived results for the abundances of the 
$\alpha$-elements, carbon, and
nitrogen. Overall, we find that ASPCAP achieves reasonably
accurate results for temperature and metallicity, but suffers from
systematic errors in surface gravity.  
We derive calibration relations that bring the raw ASPCAP results
into better agreement with independently determined stellar parameters.
The internal scatter of ASPCAP parameters within clusters suggests that, 
metallicities are measured with a precision better than 0.1~dex,
effective temperatures better than 150~K, and surface gravities better
than 0.2 dex. The understanding provided by the clusters 
and \textit{Kepler} giants on the current accuracy and precision
will be invaluable for future improvements of the pipeline.

\end{abstract}

\section{Introduction}

The study of the formation of the Milky Way Galaxy is entering a
new era, with the advent of very large surveys of the kinematics
and chemical compositions of Galactic stellar populations with
sample sizes ranging from $10^4$ to $10^7$ stars, such as RAVE
\citep{ste06}, BRAVA \citep{kun12}, SEGUE \citep{yan09}, LAMOST
\citep{zha06}, Gaia \citep{per01, lindegren01}, Gaia-ESO
\citep{gilmore01}, ARGOS \citep{nes12a,nes12b}, 4MOST \citep{dej12},
GALAH \citep{freeman01}, and 
WEAVE \citep{dalton01}. Ten years from today,
our picture of the Galaxy we live in, and with it our understanding
of its formation, will be profoundly influenced, if not radically
revised, by these major observational projects.

These various surveys cover different stellar populations within
different regions of the Galaxy. They span a range in resolution and 
spectral coverage, and they yield chemical
compositions with different precisions and accuracies. Patching
together the various pieces of this large mosaic to compose a unified
view of the Galaxy, the ultimate goal motivating all these surveys,
will be a complex and challenging endeavor. Understanding all
sources of uncertainties, both random and systematic, of the chemical
abundances delivered by each of these surveys will be key for the
success of this enterprise.

In contrast to these optical surveys, the Apache Point Observatory
Galactic Evolution Experiment \citep[APOGEE,][]{maj13, eis11} stands
out due to its focus on collecting high-resolution H$-$band data for
$10^5$ giant stars across all stellar populations. Because it
observes these stars in the near infrared (NIR), APOGEE is able to explore
remote, dust obscured, regions of the Galaxy that are beyond the
reach of optical surveys. By obtaining high
resolution (R $\sim$ 22,500) spectra for $10^5$ stars, APOGEE will 
estimate accurate abundances for up to 15 elements (depending on
temperature and metallicity) for a very large sample, and potentially
allow a statistically robust assessment of the star formation
and chemical enrichment history of all subcomponents of the Galaxy. 

APOGEE is one of four Sloan Digital Sky Survey III (SDSS-III)
experiments \citep{eis11, aihara01} and is based on the first
multi-fiber, high-resolution, NIR spectrograph ever built. This instrument 
is deployed on the SDSS 2.5m telescope at Apache Point Observatory 
\citep{gunn01}. The APOGEE spectrograph obtains 300 spectra with a 
resolution of 22,500 on three Hawaii-2RG detectors oriented to cover
the spectral interval from 15,090 to 16,990 \AA. The telescope focal
plane uses standard SDSS plug-plates, where 300 fibers with diameters 
of 2$\arcsec$ on the sky are placed on targets within a 3 degree field 
of view. The overall goal
of S/N $>$ 100 per pixel can be reached over a 3 hr integration on
targets with H=12.2. The stability of the spectrograph makes it 
possible to measure radial velocities to a precision
better than 100 m/s.  For more details on the spectrograph design
and performance, see \citet{wil10}.

Commissioning of the APOGEE spectrograph began in May 2011, and
survey operations started in September of the same year. The first
public release of APOGEE data took place as part of the tenth SDSS
data release \citep[DR10,][]{ahn01}, and includes spectra of all stars observed by
July 2012. Altogether, 180,000 spectra of nearly 60,000 stars within
170 fields were released in DR10. This release also includes spectra and derived atmospheric
parameters (including $\alpha$-elements, carbon, and nitrogen
abundances); the abundances of other elements will be made
available in a future data release. SDSS-III survey operations will 
conclude in the Summer of 2014.

The APOGEE survey is exploring all stellar components of the Milky Way
Galaxy, with particular focus on the previously less well-studied 
low latitude regions, including the dust-obscured parts of the Galactic 
disk and bulge that are accessible from  Apache Point Observatory 
\citep[APO,][]{gunn01}. APOGEE targets are selected from the 2MASS point 
source catalog \citep{cut03} with various magnitude limits, using a 
de-reddened color criterion [$(J-K)_0 > 0.5$], to minimize contamination 
of the sample by foreground dwarf and subgiant stars, and to ensure sufficiently
cool surface temperatures so that abundances can be determined. In most cases, 
final integrations are achieved through 3 to 24
separate approximately one hr visits.
Final exposure times are determined to reach 
the S/N $>$ 100/pixel requirement. More details on target 
selection are given in \citet{zasowski01}.

While the focus on NIR spectra of giants makes APOGEE unique
among all other surveys of Galactic stellar populations, it presents
some challenges for the determination of stellar parameters and
abundances and inter-comparing these to results from other surveys.
Compared to the optical, the H-band has been relatively unexplored
for detailed elemental abundance work, so that the systematic effects
intrinsic to standard abundance analyses of high-resolution NIR
spectra have not yet been subject to a thorough assessment 
\citep[but see, e.g.,][]{mel01,ori02,cun06,ryde01}. 
Additionally, most other surveys
are focused on chemical composition studies of dwarf stars. Therefore,
sample overlap between APOGEE and other surveys will be small,
making inter-survey zero-point conversions uncertain, and thus
requiring a careful alternative mapping of all sources of 
systematic effects in APOGEE parameters.

In this paper, we present an initial study of the accuracy and
precision of the primary APOGEE stellar parameters as 
available in DR10. This is achieved
for the basic stellar parameters (metallicity, surface gravity, and
temperature) using APOGEE observations of well-studied globular and
open clusters that span the parameter space of interest to APOGEE. 
Stars in clusters
are particularly important in this context due to the profusion in
the literature of high quality abundance work dedicated to giant
stars in both open and globular clusters \citep{gra04}. Moreover,
there is a steadily growing database of chemical compositions of
main sequence stars in clusters, which can in the future be used
 for understanding the impact of mixing on the abundances
of certain elements during post-main sequence evolution.  Another
important source of fundamental data for calibration of ASPCAP
output is the growing asteroseismic data base made possible by the
\textit{Kepler} satellite \citep{bo10}. Asteroseismic data are used
as an independent check on ASPCAP's surface gravities (see Section
\ref{sect:logg}), and provide broad confirmation of the trends identified in
the cluster comparisons.  Asteroseismic data also provide
methods that can be used to test the theoretical isochrones
themselves over a range of ages and chemical compositions.  The
focus of this paper, however, is on the detailed comparison between
ASPCAP parameters and those obtained by other groups for giant stars
in well known Galactic clusters and from asteroseismic surface
gravities for giants in the \textit{Kepler} fields.  

This paper is organized as follows: in Section 2 we briefly
describe the APOGEE data, the reduction pipeline, 
the APOGEE Stellar Parameters and Chemical Abundances Pipeline, 
and the Kepler asteroseismic sample. In Section 3 we present a
comparison of ASPCAP results with data from the literature for stars in
common with APOGEE and derive relations to calibrate one set of parameters 
into the other. We derived calibration relations to bring the 
ASPCAP parameters into agreement with independent parameter estimates.
Finally, we summarize our results in Section 4.

\section{Data Analysis}

\subsection{Data Reduction}

The APOGEE observations in a given field consist of multiple exposures, each
of which yields multiple non-destructive readouts of
the NIR detectors via the up-the-ramp-sampling method. 
Because the spectra are slightly undersampled
at the short wavelength end of the spectrum, exposures are taken
in pairs with the detectors physically shifted in the spectral direction
by 0.5 pixels between exposures. The raw data are reduced to 
well-sampled 1D spectra using a custom pipeline.

The pipeline uses the up-the-ramp data cubes 
and calibration data (darks and flats) to construct a calibrated
2D image for each exposure, extracts the 300 spectra from these 2D
images, applies a wavelength calibration, subtracts sky and performs a
telluric correction using data from sky fibers and fibers placed
on hot stars. It also assembles well-sampled spectra from the dither pairs,
and applies flux calibration. Data from multiple visits
to the same field on different nights are combined after determination
of the relative RVs of the different observations.  Along with the
final spectra, the pipeline produces error and mask arrays, RVs of
the individual visits, and a number of intermediate data products.
Spectra from the 3 APOGEE detectors are combined together into one file, but 
continuum-normalized separately from each other. 

Calibration lamps and sky lines are used to obtain measurements of
the line spread function. These yield a typical FWHM resolving power of $\sim$22,500,
although there are some variations, at the $\sim$10\% level, with location
on the chip and with wavelength.
More details about the data reduction pipeline and the instrument
performance will be given by \citet{nid13}.

\subsection{ASPCAP and FERRE}

The outputs from the data reduction pipeline are fed to the APOGEE
Stellar Parameters and Chemical Abundances Pipeline (ASPCAP). 
This pipeline already uses simultaneously the spectra from the 
3 detectors. Details
of pipeline operation and performance tests will be described in a 
future paper, but we summarize the basic principles here.

ASPCAP searches a precomputed grid of synthetic spectra for the
combination of atmospheric parameters consisting of effective
temperature (T$_{\rm eff}$), surface gravity ($\log~g$), metallicity
([M/H]), carbon ([C/M]), nitrogen ([N/M]), and  $\alpha$ ([$\alpha$/M])
abundances associated with a synthetic spectrum that best reproduces
the observed pseudo-continuum-normalized fluxes. 

The abundance of each individual element X heavier than helium, is defined as

\begin{equation}
[X/H] = \log_{10}(n_X/n_H)_{\rm star} - \log_{10}(n_X/n_H)_\odot 
\end{equation}

where $n_X$ and $n_H$ are respectively the number of atoms of
element X and hydrogen, per unit volume in the stellar photosphere.
We define [M/H] as an overall scaling of metal abundances
with a solar abundance pattern, and [X/M] as 
the deviation of element X from the solar abundance pattern:

\begin{equation}
[X/M] = [X/H] - [M/H]
\label{eqn:mh}
\end{equation}

In the current version of ASPCAP, [C/M], [N/M], [$\alpha$/M] 
are allowed 
to vary because these elements are seen to depart from
the solar abundance pattern and also because a substantial part of
the line opacity in the H band is sensitive to those abundances,
particularly due to the presence of thousands of lines from 
CN, OH, and CO molecules. The $\alpha$ elements considered in APOGEE are 
the following: O, Ne, Mg, Si, S, Ca, and Ti. 
For all other elements, we presently set [X/M] = 0, but intend to
extend the analysis to additional species in future releases. 
Additional explanations of our metallicity definition can be found in Section 3.2.

ASPCAP has two main parts. At its core, the FORTRAN90 code FERRE 
performs the parameter search by evaluating the differences between
 observed and model fluxes by interpolation on a grid of
pre-computed spectra. The remainder of the pipeline,
which we refer to as the IDL wrapper, reads and prepares the reduced
APOGEE spectra for FERRE, executes the code, and collects and writes
the output into FITS tables.  For each input spectrum, a first
pass derives the main atmospheric parameters mentioned above, and
a second pass determines individual chemical abundances, although
only the first is currently operational. The long-range goal is
to exploit the rich chemical information content in each APOGEE
spectrum to derive abundances for approximately 15 elements, but
presently only the outputs of the first pass, namely, overall
metallicity and the abundances of $\alpha$ elements, carbon, and
nitrogen, are being determined.

FERRE has evolved from the code used by \citet{allende01}, 
and implements a number of optimization and interpolation algorithms.
The merit function for the optimization is a straight $\chi^2$
criterion. The grids of model fluxes on which the code interpolates
are stored in memory or in a database, with the former solution
being typically faster. With about 10$^4$ wavelength bins represented in the
order of 10$^6$ models, the model grids used by ASPCAP are
massive.

FERRE performs 12 searches for each input spectrum.  These are
initialized at the center of the grid for [C/M], [N/M] and [$\alpha$/M],
at two different places symmetrically located from the grid center
for [M/H] and $\log~g$, and at three for T$_{\rm eff}$. Random
contributions to the error bars are calculated by inverting the
curvature matrix following the discussion in \citet{press01}. FERRE
is parallelized using OPENMP\footnote{http://openmp.org/}.

ASPCAP fits model spectra to observations in normalized
flux. The IDL wrapper normalizes independently the
sections of the reduced spectra falling on each of the three
detectors. Because for the cool stars that APOGEE mainly 
observes the continuum is uncertain, ASPCAP relies on applying the same 
continuum normalization procedure to both model and observed spectra. 
The code iteratively fits a polynomial and sigma-clips points above 
and below thresholds to the spectra, observed and model, in the same 
fashion that it is traditionally done for observations. 
The three spectra are combined after this normalization.  

ASPCAP groups the observed stars according to
four broad spectral classes, each with an associated
model library. The purpose of splitting the libraries by effective temperature 
is simply to keep the size of grids manageable. 
The wrapper manages the execution of FERRE jobs for
each class, and combines the output by selecting the best solution for
each spectrum processed through multiple searches.
Finally, the resulting parameters, error covariance
matrices, best-fitting spectra, and other relevant quantities are
stored in compact FITS files.

\subsection{Spectral Synthesis and Linelist}

The libraries of model stellar spectra currently in use have been
calculated using Kurucz model atmospheres \citep{castelli01},
up-to-date continuum \citep{allende02, allende03} and line opacities, 
and the spectral synthesis code ASS$\epsilon$T
\citep{koesterke01, koesterke02}.  All model atmospheres were
constructed with a constant value for the microturbulence of 2.0
km s$^{-1}$, and solar abundance ratios scaled to the metallicity
for all metals. In contrast, a wide range of microturbulence values
and abundance ratios for [C/M], [N/M] and [$\alpha$/M] were adopted
for the spectral synthesis calculations. However, for DR10 results,
a relation between micro-turbulent velocity and surface gravity 
($v_{micro} = 2.24 - 0.3 \times \log g$) was found to describe
well the results from fitting with 7 parameters (the usual six and 
microturbulence), and adopted for subsequent work. Future plans 
are to use consistent atmospheres with non-solar abundance ratios 
to match those used for the spectral synthesis based on model 
atmospheres calculated by \citet{meszaros01}.

The line list adopted for the ASPCAP analysis includes both
atomic and molecular species.  The molecular line list, compiled 
from literature sources, included: CO, OH, CN, C$_2$, H$_2$, and SiH. 

All of the molecular data were adopted from the literature
without modifications with the exception of a few obvious typographical
corrections. The original atomic line list was compiled from
a number of literature sources and includes theoretical,
astrophysical, and laboratory oscillator strength values. Once we
had what we considered to be our best literature atomic values, we
allowed the transition wavelengths, oscillator strengths, and damping
constants to vary to fit to the solar spectrum.
For this purpose we used a customized version of the LTE spectral
synthesis code MOOG \citep{sneden06}. 

For this work, two libraries were used: one spanning
span spectral types from early-M to K-type stars ($3500 \le T_{\rm eff} \le 5000$~K), 
and the other covering G and early-F type stars ($4750 \le T_{\rm eff} \le 6500$~K).

\subsection{Cluster calibration targets}

APOGEE observed 20 open and globular clusters during the first year
of survey operations, partly for calibration purposes. The
calibration clusters were selected to span a wide range of metallicities
and also to have well-measured abundances from previous studies.
Table 1 lists these calibration clusters, along with the adopted
[M/H], E($B-V$), and age from the literature.  For the open and
globular clusters, targets were selected as cluster members if 1)
there was published abundance information on the star as a cluster
member, 2) they were radial velocity members, or 3) if they had a
probability $>$50\%  of being a cluster member based on their proper
motions.  For stars with existing abundance measurements and
atmospheric parameters, the references are provided in Table 1. The
cluster target selection is described in more detail  by
\citet{zasowski01}.

\begin{deluxetable}{llrcrcrr}
\tabletypesize{\scriptsize}
\tablecaption{Properties of Clusters from the Literature}
\tablewidth{0pt}
\tablehead{
\colhead{ID} & \colhead{Name} & 
\colhead{[Fe/H]} & \colhead{Ref.\tablenotemark{a}} & 
\colhead{E(B$-$V)} & \colhead{Ref.\tablenotemark{b}} & 
\colhead{log age\tablenotemark{c}} & \colhead{Individual Star Ref. \tablenotemark{d}} }
\startdata
NGC 6341	& M92		& -2.35$\pm$0.05 & 1 & 0.02 & 1 & 10.0 & 9, 10\\
NGC 7078	& M15		& -2.33$\pm$0.02 & 1 & 0.10 & 1 & 10.0 & 9, 31, 32, 33, 34, 35\\
NGC 5024	& M53		& -2.06$\pm$0.09 & 1 & 0.02 & 1 & 10.0 & \\
NGC 5466	& 		& -1.98$\pm$0.09 & 1 & 0.00 & 1 & 10.0 & 25\\
NGC 4147	& 		& -1.78$\pm$0.08 & 1 & 0.02 & 1 & 10.0 & \\
NGC 7089	& M2		& -1.66$\pm$0.07 & 1 & 0.06 & 1 & 10.0 & \\
NGC 6205	& M13		& -1.58$\pm$0.04 & 1 & 0.02 & 1 & 10.0 & 1, 2, 28\\
NGC 5272	& M3		& -1.50$\pm$0.05 & 1 & 0.01 & 1 & 10.0 & 1, 2, 28, 29, 30\\
NGC 5904	& M5		& -1.33$\pm$0.02 & 1 & 0.03 & 1 & 10.0 & 18, 19, 20, 21, 22, 23\\
NGC 6171	& M107  	& -1.03$\pm$0.02 & 1 & 0.33 & 1 & 10.0 & 26, 27\\
NGC 6838	& M71		& -0.82$\pm$0.02 & 1 & 0.25 & 1 & 10.0 & 11, 12, 13, 14, 15\\
NGC 2158	& 		& -0.28$\pm$0.05 & 2 & 0.43 & 2 & 9.0 & \\
NGC 2168	& M35		& -0.21$\pm$0.10 & 5 & 0.26 & 5 & 8.0 & \\
NGC 2420	& 	  	& -0.20$\pm$0.06 & 2 & 0.05 & 2 & 9.0 & 6, 7, 8\\
NGC 188		& 	  	& -0.03$\pm$0.04 & 2 & 0.09 & 2 & 9.6 & 8 \\
NGC 2682	& M67		& -0.01$\pm$0.05 & 2 & 0.04 & 2 & 9.4 & 7, 16, 17 \\
NGC 7789	& 	  	& +0.02$\pm$0.04 & 2 & 0.28 & 2 & 9.2 & 8, 24\\
M45		& Pleiades	& +0.03$\pm$0.02 & 3 & 0.03 & 5 & 8.1 & \\
NGC 6819	& 	  	& +0.09$\pm$0.03 & 6 & 0.14 & 3 & 9.2 & \\
NGC 6791	& 	  	& +0.47$\pm$0.07 & 4 & 0.12 & 4 & 9.6 & 3, 4, 5\\
\enddata
\tablenotetext{a}{[Fe/H] references: (1) \citet{carretta01}; (2) \citet{jacobson01}; (3) \citet{soderblom01}; 
(4) \citet{carretta02}; (5) \citet{barrado01}; (6) \citet{bragaglia01}}
\tablenotetext{b}{E(B$-$V) references: (1) \citet{harris01}; (2) \citet{jacobson01}; (3) \citet{bragaglia01}; 
(4) \citet{carretta02}; (5) http://www.univie.ac.at/webda/}
\tablenotetext{c}{Ages used in isochrones, open clusters: http://www.univie.ac.at/webda/}
\tablenotetext{d}{Individual star references: (1) \citet{sneden01}; (2) \citet{cohen01}; (3) \citet{origlia01};
(4) \citet{carraro01}; (5) \citet{carretta02}; (6) \citet{friel01}; (7) \citet{pancino01}; (8) \citet{jacobson01};
(9) \citet{sneden02}; (10) \citet{roederer01}; (11) \citet{mel02}; (12) \citet{briley01}; (13) \citet{shetrone01}; 
(14) \citet{lee01}; (15) \citet{yong01}; (16) \citet{taut01}; (17) \citet{jacobson01}; (18) \citet{lai01};
(19) \citet{ivans01}; (20) \citet{koch01}; (21) \citet{sneden03}; (22) \citet{ram01}; (23) \citet{yong02};
(24) \citet{taut02}; (25) \citet{shetrone01}; (26) \citet{connell01}; (27) \citet{carretta01}; (28) \citet{cavallo01};
(29) \citet{kraft01}; (30) \citet{kraft02}; (31) \citet{minniti01}; (32) \citet{otsuki01}; (33) \citet{sneden04};
(34) \citet{sneden05}; (35) \citet{sobeck01}}
%\tablecomments{}
\end{deluxetable}

The APOGEE observations and ASPCAP analysis were used to refine the
list of cluster members. Cluster membership for most of the stars in our sample was
established in the original works from which we adopted the literature
values. In fact, almost all radial velocities differed from
APOGEE cluster averages by less than 15$-$20~km/s. Only a few outliers with 
differences of 30~km/s or
higher were excluded from the sample, because of possible binarity or 
probable non-membership. In addition, we 
removed any stars that had a significantly different metallicity (usually
$>$0.3~dex, or about 3 $\sigma$)  from the ASPCAP average,
or if the position of the star on the HR diagram was far from
the RGB. Based on these criteria, about 90\% of the originally
selected stars were adopted as cluster members.

For the discussion on the accuracy and precision of the derived
parameters, we restrict our analysis to stars with S/N $>$ 70 (where
S/N is determined per pixel in the final combined spectrum, which
has $\sim$ 3 pixels per resolution element), as tests carried out
with ASPCAP indicate that this value is the minimum  required
to derive reliable stellar parameters.

The analysis was also restricted to giant stars with $\log~g$ $<$ 3.5,
because most of the stars observed in the calibration of clusters
are evolved; we stress that {\it all} calibrations presented
in Section 3 apply only to stars with $\log~g$ $<$ 3.5. 
Dwarfs were observed in only two clusters, M35 and the
Pleiades; therefore, because both of these clusters have near-solar 
metallicity, there is insufficient information to determine the 
accuracy of ASPCAP parameters for dwarfs outside the solar metallicity regime.
We note that the restriction to evolved stars is not overly limiting
from the point of view of calibrating the survey, as giants are the
main targets and expected (and observed) to comprise $\sim$ 80\%
of the entire APOGEE sample.

\subsection{Standard Stars}

In addition to the APOGEE observations of cluster stars, we also
tested the results of ASPCAP and the adopted line list by analyzing
four bright, well-studied stars that have high resolution (varied
from R = 45,000 to 100,000), near-IR spectra obtained via the Kitt
Peak National Observatory Fourier transform spectrometer on the
Mayall 4-m telescope \citep{smith01}. This red giant sample was
chosen to cover a large part of ASPCAP's expected stellar parameter
range.  

The parameters were derived from fundamental measurements in the 
case of $T_{\rm eff}$ and $\log g$ (angular diameters and parallaxes), 
and spectral synthesis in the case of the abundances, using the line
list compiled especially for APOGEE (see Section 2.4). The
selected stars are all nearby red giants with well-known stellar
parameters, and provide a more direct check than other spectroscopic
studies, because the spectral synthesis adopted 
the same line list and model atmosphere grid as used by
ASPCAP, although the spectra were obtained with a different
spectrograph and are higher resolution than APOGEE spectra.

The discussion of the differences between the manually derived
parameters and ASPCAP values are detailed in the relevant physical
parameters section in the Discussion. The list of stars and their
atmospheric parameters from both ASPCAP and \citet{smith01} are
listed in Table 2.

\begin{deluxetable}{lrrrrrr}
\tabletypesize{\scriptsize}
\tablecaption{Properties of Standard Stars from ASPCAP and from Manual Analysis}
\tablewidth{0pt}
\tablehead{
\colhead{Star} & 
\colhead{$T_{\rm eff}$\tablenotemark{a}} & \colhead{T$_{\rm eff}$\tablenotemark{b}} &
\colhead{[M/H]\tablenotemark{a}} & \colhead{[M/H]\tablenotemark{b}} & 
\colhead{$\log~g$\tablenotemark{a}} & \colhead{$\log~g$\tablenotemark{b}}}
\startdata
$\alpha$ Boo & 4158 & 4275 & -0.61 & -0.47 & 2.10 & 1.70 \\
$\beta$ And  & 3739 & 3825 & -0.41 & -0.22 & 1.40 & 0.90 \\
$\delta$ Oph & 3753 & 3850 & -0.15 & -0.01 & 1.52 & 1.20 \\
$\mu$ Leo  & 4476 & 4550 & 0.25 & 0.31 & 2.87 & 2.10 \\
\enddata
\tablenotetext{a}{Parameters derived by ASPCAP}
\tablenotetext{b}{Parameters derived by \citet{smith01}}
\end{deluxetable}

\subsection{Red Giants in the \textit{Kepler} Field}

Asteroseismology provides a way of determining the surface
gravities of a star that is essentially independent from a spectroscopic
analysis. The only non-seismic information required is the effective
temperature of the star and of the Sun as well as the solar value of 
$\log~g$. There are 673 giants that the Kepler Asteroseismic Science 
Consortium (KASC) identified as members of the
``gold sample". The gold sample consists of stars observed nearly 
continuously for 600 days \citep[Kepler run Q1-Q7,][]{hekker02}. 
For this sample robust seismic parameters have been derived using 
different methods \citep{hekker02, kallinger01, mosser02}. APOGEE observed 280 of those stars 
in the DR10 period. These have seismic surface gravities derived using 
APOGEE temperatures to test gravities derived by ASPCAP. This sample provides
an independent way from isochrones and spectroscopic gravities of 
checking the $\log~g$ determined by ASPCAP. Additionally, the quoted 
uncertainties of the asteroseismic log g are often an order of 
magnitude lower than those quoted in spectroscopic analyses.

%Several studies have explored the accuracy of asteroseismic $\log~g$
%values.  
For main-sequence and subgiant stars the accuracy of the
asteroseismically determined $\log~g$ has been investigated by
comparisons with $\log~g$ values from classical spectroscopic methods
\citep[e.g.;][]{morel2012} and from independent determinations of
radius and mass \citep[e.g.;][]{creevey2012,creevey2013}. These
studies found good agreement between the gravities inferred from
asteroseismology and spectroscopy.
%, which supports the use of
%asteroseismic $\log~g$ values. 
For more evolved stars a small sample has
been investigated by \citet{morel2012} who found good agreement between 
asteroseismic $\log~g$ and gravities derived using classical methods for
$\log~g$ values down to 2.5~dex. \citet{Thygesen2012} showed a
comparison between spectroscopic and asteroseismic $\log~g$ values
for 81 low-metallicity stars with $\log~g$ down to 1.0~dex.  This
sample also revealed good agreement with previous studies. A much
larger sample was explored by \citet{hekker2010} and the results from
this work provide the basis for the analysis performed here.

Although surface gravity can be computed directly from asteroseismic 
scaling relations, more precise and reliable results are 
obtained by using grid-based modeling \citep{gai2011}. 
The grid-based modeling used in \citet{hekker02} is performed by two 
independent 
implementations based on the recipe described by \citet{basu01}. One implementation uses BaSTI models 
\citep{cassisi2006}. The other implementation uses YY isochrones \citep{demarque2004}, models constructed 
with the Dartmouth Stellar Evolution Code \citep{dotter2007} and the model grid of \citet{marigo2008}.
All implementations provided consistent results. 

\citet{gai2011} previously showed that asteroseismic $\log~g$ is largely model independent and 
this is confirmed in \citet{hekker02}. \citet{gai2011} showed that an asteroseismic 
$\log~g$ can be obtained precisely and accurately with both direct and grid-based methods.

\section{Discussion}

\begin{deluxetable}{lrrrrrrrrrrrrrrrrr}
\rotate{}
\tabletypesize{\scriptsize}
\tablecaption{Properties of Stars Used for Validation of ASPCAP}
\tablewidth{-210pt}
\tablehead{
\colhead{2MASS ID} & 
\colhead{Cluster} & 
\colhead{v$_{\rm helio}$} &
\colhead{T$_{\rm eff}$} & 
\colhead{T$_{\rm eff}$} & 
\colhead{$\log~g$} & 
\colhead{$\log~g$} & 
\colhead{[M/H]} & 
\colhead{[M/H]} & 
\colhead{[C/M]} & 
\colhead{[N/M]} & 
\colhead{[$\alpha$/M]} & 
\colhead{S/N} & 
\colhead{J\tablenotemark{a}} & 
\colhead{H\tablenotemark{a}} & 
\colhead{K\tablenotemark{a}} & 
\colhead{T$_{\rm eff}$ $\sigma$} & 
\colhead{[M/H] $\sigma$}  \\
\colhead{} & 
\colhead{} & 
\colhead{(km/s)} &
\colhead{ASP.} & 
\colhead{cor.} & 
\colhead{ASP.} & 
\colhead{cor.} & 
\colhead{ASP.} & 
\colhead{cor.} & 
\colhead{ASP.} & 
\colhead{ASP.} & 
\colhead{ASP.} & 
\colhead{} & 
\colhead{} & 
\colhead{} & 
\colhead{} & 
\colhead{cor.} & 
\colhead{cor.}
}
\startdata
2M17162228+4258036 	&M92 	&-118.06 &5067.8 	&4995.2 	&2.56 	&2.08 	
&-1.94 	&-2.20 	&0.58 	&0.90 	&0.04 	&163.1 	&12.826  &12.267 	&12.247 	&171.3 	&0.134\\
2M17163427+4307363 	&M92 	&-119.60 &4776.2 	&4819.3 	&1.85 	&1.35 	
&-2.13 	&-2.32 	&0.19 	&0.78 	&0.15 	&142.6 	&11.948  &11.407 	&11.340 	&176.3 	&0.139\\
2M17163577+4256392 	&M92 	&-117.42 &5200.8 	&5075.4 	&2.87 	&2.40 	
&-1.92 	&-2.21  &0.83 	&0.94 	&0.06 	&125.0 	&13.171  &12.675 	&12.612 	&171.7 	&0.135\\
2M17164330+4304161 	&M92 	&-121.02 &5200.6 	&5075.3 	&2.91 	&2.43 	
&-1.95 	&-2.24 	&0.86 	&0.95 	&0.13 	&132.2 	&13.091  &12.618 	&12.504 	&172.9 	&0.136\\
2M17165035+4305531 	&M92 	&-117.37 &4948.4 	&4923.2 	&2.14 	&1.65 	
&-2.11 	&-2.34  &0.44 	&0.91 	&0.17 	&160.0  &12.446  &11.969 	&11.908  &176.9 	&0.139\\
\enddata
\tablecomments{Notations: ASP.: ASPCAP raw parameters, cor.: corrected parameters. 
This table is available in its entirety in machine-readable form in the online journal. A portion 
is shown here for guidance regarding its form and content. After DR10 was published we discovered that 
four stars had double entries with identical numbers in this table (those are deleted from this table, 
thus providing 559 stars). All calibration equations were derived with those four double entries in 
our tables, but because DR10 is already published we decided not to change the fitting equations 
in this paper. This problem does not affect the effective temperature correction. 
The changes in the other fitting equations are completely negligible and have no affect in any scientific application. 
The parameters published in DR10 are off by $<1$~K in case of the effective temperature error correction, 
and by $<$0.001~dex for the metallicity, metallicity error, and surface gravity correction.}
\tablenotetext{a}{J, H, K photometry is taken from the 2MASS catalog \citep{struskie01}. }
%\tablenotetext{b}{Before the metallicity correction}
%\tablenotetext{c}{After the metallicity correction}
\end{deluxetable}

In this section, we focus on systematic and random errors for the most important ASPCAP parameters
(effective temperatures, metallicities, and surface gravities). We also include some 
discussion of [$\alpha$/M], [C/M],and [N/M], but this is limited due to the lack of corresponding data for many
stars. Since the ASPCAP fits are based on model atmospheres and synthetic spectra
that are necessarily imperfect, systematic offsets in derived stellar parameters are
not totally unexpected. To account for this, 
the SDSS Data Release includes not only the raw ASPCAP results, but also
``calibrated'' values in which we apply offsets to the raw ASPCAP results. This
section presents the derivation of the calibration relations used for DR10 data,
based on a comparison of ASPCAP results for objects with independently determined parameters.
We use the scatter around the calibration relations to provide an estimate
of the precision of the calibrated ASPCAP results.

We compare with spectroscopic, photometric, and asteroseismic
diagnostics.  In general, our approach is to use comparisons with
the ensemble of cluster and asteroseismic data to measure the
presence of systematic trends.  Since most of APOGEE's targets
have $-0.5 < $[M/H]$ < 0.1$, our emphasis is on the calibration for
this particularly important metallicity region. We use the scatter in these
calibrated results and those within clusters to provide an estimate
of the precision of the calibrated ASPCAP results. The list of 559 stars 
used in this analysis, and their original and corrected ASPCAP values are listed in Table 3. 

\subsection{Accuracy of Effective Temperature}

In order to test the accuracy of the ASPCAP effective temperatures,
we derived photometric temperatures from 2MASS $J-H$ and $J-K_{s}$
colors \citep{struskie01}. De-reddened $J-H$ and $J-K_{s}$ were calculated
from $E(B-V)$, listed in Table 2 for each cluster, using $E(J-H) =
0.326 \cdot E(B-V)$ \citep{schlegel01} and 
$E(J-K_{s}) = 0.46 \cdot E(B-V)$. We chose to use calibrations published by
\citet{gonzalez01}, which are based on 2MASS $J$, and $Ks$
magnitudes. We also compared these effective temperatures with color-temperature 
calibrations by \citet{alonso01, alonso02, houdashelt01}.  De-reddened
colors had to be converted from the 2MASS photometric system to the
CIT and TCS system\footnote{http://www.astro.caltech.edu/~jmc/2mass/v3/transformations/} 
for these latter calibrations.
\citet{cas01} implied that the \citet{gonzalez01} temperature scale 
may only have 30$-$40~K systematic difference from the absolute temperature scale, 
thus we chose the \citet{gonzalez01} relation as our primary calibrator.

As an independent check, and
because photometric temperatures depend on the assumed extinction, 
and weakly on metallicity, we also compare
the ASPCAP temperatures with temperatures derived from high resolution
spectroscopic studies in the literature (Table 1).

\begin{figure}[!ht]
\includegraphics[width=4.5in,angle=270]{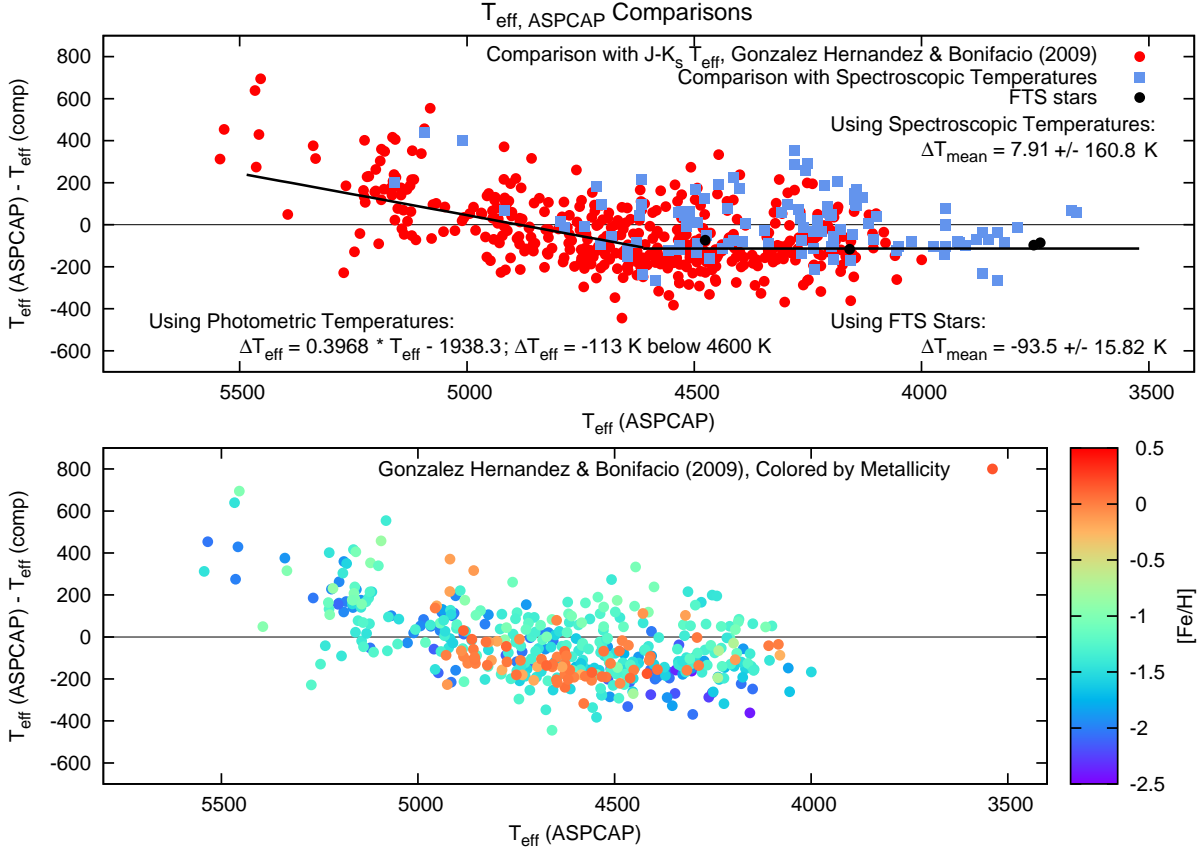}
\caption{\textit{Top panel}: ASPCAP effective temperatures compared
to photometric calibrations (red squares), spectroscopic literature
values (blue squares), and the four standard stars (black dots). The
solid line represents the empirical calibration to the photometric
temperatures. See Section 3.1 for discussion.  \textit{Bottom panel}:
ASPCAP effective temperatures compared to photometric calibrations
color coded by metallicity. Only giant stars with $\log~g$ $<$ 3.5 are plotted in both panels. }
\label{fig:teff}
\end{figure}

Figure \ref{fig:teff} shows the differences
between ASPCAP T$_{\rm eff}$ and $J-K_{s}$-based T$_{\rm eff}$ and
those from other high-resolution spectroscopic studies.  The
agreement is generally good in both cases, however, there is a slight
evidence of systematic offsets.
The comparison with the photometric temperatures shows a
linear trend in the differences as a function of ASPCAP T$_{\rm
eff}$ between 4750 and 5500~K. Between 4600 and 5000~K,
photometric \citep{gonzalez01} and ASPCAP temperatures are in good agreement,
although a slight trend towards higher ASPCAP T$_{\rm eff}$ is
also visible.  Below 4600~K, the ASPCAP temperatures are cooler
than those based on $J-K_{s}$ by about 100$-$200~K.  The
photometric temperatures depend on the reddening
used, thus for high reddening clusters 
($E[B-V] > 0.1$: M107, M71, NGC 2158, NGC 7789, NGC 6819, NGC
6791) errors in the reddening may lead to measurable linear shifts
in the differences. An error of $\pm$10\% in the reddening introduces 
an error of about $\pm$50~K in temperature, which 
is not enough to explain the differences between ASPCAP and the 
photometric scale by \citet{gonzalez01}, because our tests with zero reddening show that 
similar offsets remain. An additional error source is the zero point
used in the calibration to the fundamental temperature scale, and
we found that there is an average difference of 70~K between  
the calibrations of \citet{gonzalez01} and the combined calibrations of \citet{alonso01,
alonso02, houdashelt01}.

Despite these uncertainties we provide a
calibration that ties ASPCAP temperatures to the scale used by
\citet{gonzalez01} between 4600~K and 5500~K:

\begin{eqnarray}
T_{corrected} &= T_A - 0.3968 \ T_{A} + 1938.3 &; 4600 < T < 5500 \\
 &= T_A + 113.3  &;  3500 < T < 4600 \nonumber
\label{eqn:teff}
\end{eqnarray}
where $T_A$ is the raw ASPCAP effective temperature. The equation is 
only valid for stars with $\log~g$ $<$ 3.5. Note that both the original
spectroscopic $T_{\rm eff}$s and those calibrated with this relationship
are provided in DR10.

The bottom panel of Figure \ref{fig:teff} shows the metallicity dependence of
the photometric scale comparison by color-coding the points according to metallicity.
Below 5000~K both metal-poor and metal-rich stars are present, 
but no strong metallicity dependence can be perceived in the data.  
Above 5000~K the clusters
presented in this paper only contain metal-poor stars, but expanding
the comparison to field stars observed by APOGEE shows no metallicity
dependence in this temperature range.
However, because both metal-poor and
metal-rich stars show large, 200$-$500~K differences above 5000~K,
we believe that ASPCAP temperatures are overestimated in
this region.

The comparison of the raw ASPCAP temperatures with the literature spectroscopic 
temperatures shows very good
agreement for $T_{\rm eff} < 5000$~K. The average difference is only
8~K, while the scatter is 161~K, of which a significant component could 
come from uncertainties in the literature values.
The good agreement with spectroscopic gravities suggests that
no empirical calibrations are required below 5000~K, which is in
mild conflict with the photometric temperature comparison.

Comparing ASPCAP values with the fundamental (non-spectroscopic) temperatures  
from \cite{smith01} for the
four standard stars, we find the latter to be warmer by an average
of 94 $\pm 9$ K.  This small discrepancy induces a metallicity difference 
between the two analyses of $-0.13 \pm 0.03$ dex (see Section
3.2).  These stars have well defined parameters, and the systematic
difference of temperatures from ASPCAP agrees well with the photometric
calibrations.  We note that these stars were not observed by APOGEE,
thus a simplified version of ASPCAP had to be used, which may
introduce systematic differences, even if the spectral synthesis was
based on the same line list and model atmosphere grid as for
APOGEE.

In the end, we recommend using the corrected temperatures (equation 3) 
that bring the ASPCAP temperatures to the IRFM photometric scale. 
The advantage of doing so is that it is tied to the 
fundamental temperature scale, especially at solar abundance. 
Inspection of Figure \ref{fig:teff} reveals a statistically significant 
but modest ($\sim$100~K) difference between the spectroscopic and 
fundamental scales. 
The precision of ASPCAP T$_{\rm eff}$ is discussed in more detail in Section 3.5.

\subsection{Accuracy of Metallicity}

\begin{deluxetable}{lrrrrrrr}
\tabletypesize{\scriptsize}
\tablecaption{Properties of Clusters Derived from ASPCAP}
\tablewidth{0pt}
\tablehead{
\colhead{ID} & \colhead{Name} & 
\colhead{N\tablenotemark{a}} &
\colhead{[M/H]\tablenotemark{b}} & \colhead{[M/H]\tablenotemark{c}} & 
\colhead{[M/H] rms\tablenotemark{b}} & 
\colhead{T$_{\rm eff}$ rms} & \colhead{[$\alpha$/H] rms}}
\startdata
NGC 6341       & M92	       & 48 & -2.03 & -2.26	  & 0.12 & 179.2  & 0.10 \\
NGC 7078       & M15	       & 11 & -2.11 & -2.25	 & 0.14 & 141.1 & 0.08 \\
NGC 5024       & M53	       & 16 & -1.94 & -2.06	 & 0.11 & 113.5 & 0.06 \\
NGC 5466       &	       & 8 &  -1.90 & -2.02	& 0.08 & 135.1 & 0.06 \\
NGC 4147       &	       & 3 &  -1.66 & -1.82	& 0.21 & 307.7 & 0.05 \\
NGC 7089       & M2	       & 19 & -1.46 & -1.58	 & 0.09 & 148.1 & 0.06 \\
NGC 6205       & M13	       & 71 & -1.43 & -1.60	 & 0.12 & 146.9  & 0.06 \\
NGC 5272       & M3	       & 73 & -1.39 & -1.55	& 0.12 & 186.7 &  0.06 \\
NGC 5904       & M5	       & 103 & -1.19 & -1.34	  & 0.12 & 183.4 & 0.06\\
NGC 6171       & M107	       & 18 & -0.92 & -1.02	 & 0.21 & 150.8 & 0.04 \\
NGC 6838       & M71	       & 7 &  -0.72 & -0.75	& 0.04 & 100.2  & 0.04 \\
NGC 2158       &	       & 10 & -0.15 & -0.17	 & 0.03 & 101.7 & 0.02 \\
NGC 2168       & M35	       & 1 & -0.11  & \nodata	& \nodata & \nodata & \nodata \\
NGC 2420       &	       & 9 & -0.20  & -0.22	& 0.04 & 64.0 & 0.03 \\
NGC 188        &	       & 5 & +0.03  & +0.06	 & 0.02 & 123.1 & 0.02 \\
NGC 2682       & M67	       & 24 & +0.0  & +0.03	& 0.06 & 64.8 & 0.03 \\
NGC 7789       &	       & 5 & -0.02  & +0.01	& 0.05 & 61.5 & 0.02 \\
M45	       & Pleiades      & 75 & -0.05 & \nodata	 & 0.18 & \nodata & 0.11 \\
NGC 6819       &	       & 30 & +0.02 & +0.05	 & 0.06 & 77.8 & 0.02 \\
NGC 6791       &	       & 23 & +0.25 & +0.37	 & 0.07 & 50.6 & 0.07 \\
\enddata
\tablenotetext{a}{N: number of stars used in the analysis}
\tablenotetext{b}{Before the metallicity correction}
\tablenotetext{c}{After the metallicity correction}
\end{deluxetable}

As any of the other primary physical parameters, metallicity
is crucial for determining the rest of the abundances. In
ASPCAP, the metallicity parameter ([M/H]) is used to track 
variations in all metal abundances  locked in solar proportions, 
while deviations from solar abundance ratios are tracked by additional
parameters, such as carbon, nitrogen and the alpha element abundances:
[C/M], [N/M], and [$\alpha$/M], respectively (see Section 2.2).
Here, we derive calibrations based on ASPCAP measurements of clusters with
known metallicity. Since the adopted cluster metallicities are
generally measurements of [Fe/H] (i.e., iron specifically),
calibrations derived from these data tie ASPCAP [M/H]
to literature values of [Fe/H]. For the most part, therefore, one can
treat the calibrated values of [M/H] as if they were [Fe/H],
and [X/M] values as [X/Fe]. Our tests in clusters also show that using
only iron lines to derive [Fe/H] reproduces the uncalibrated [M/H] values
to better than 0.1 dex. We maintain the M notation because the raw [M/H] values
are derived fitting lines of many elements, not just iron. In
following data releases we intend to publish individual
abundances for many elements, including [Fe/H] (which would
allow non-zero values of [Fe/M]), based on fittings to specific
absorption lines in the APOGEE spectral window, establishing the
detailed abundance pattern for each individual star.

\begin{figure}[!ht]
\includegraphics[width=4.5in,angle=270]{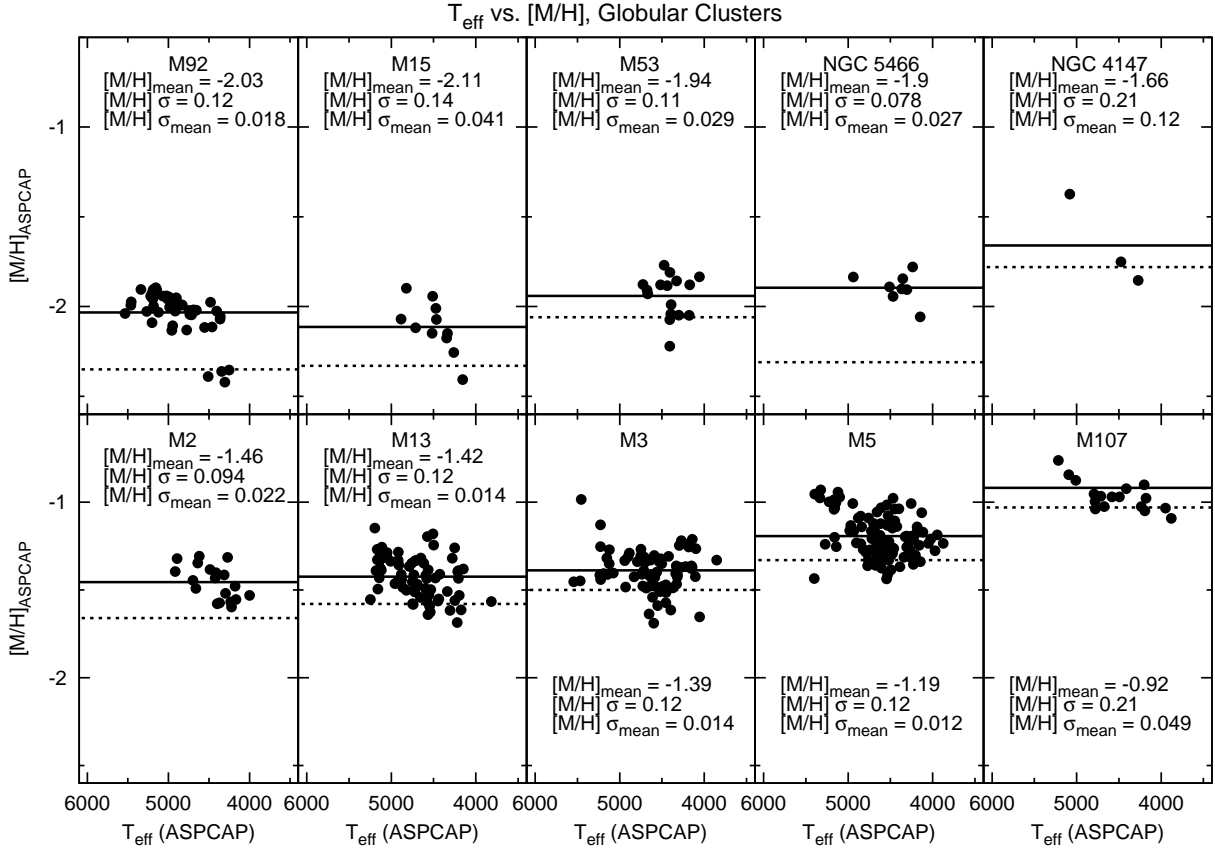}
\caption{[M/H] as a function of effective temperature derived with
ASPCAP for globular clusters. Average cluster metallicities and the
standard deviation of both the average and the distribution is
labeled. Solid line marks the
average of ASPCAP metallicities, the dashed line denotes the cluster
average from literature, listed in Table 1. ASPCAP metallicities
are 0.2$-$0.3~dex higher than the literature (Table 1.) below
[M/H]=$-$0.7. A slight trend with temperature is visible in the
data for some of the globular clusters. See Section 3.2 for discussion.
}
\label{fig:metglob}
\end{figure}

\begin{figure}[!ht]
\includegraphics[width=4.5in,angle=270]{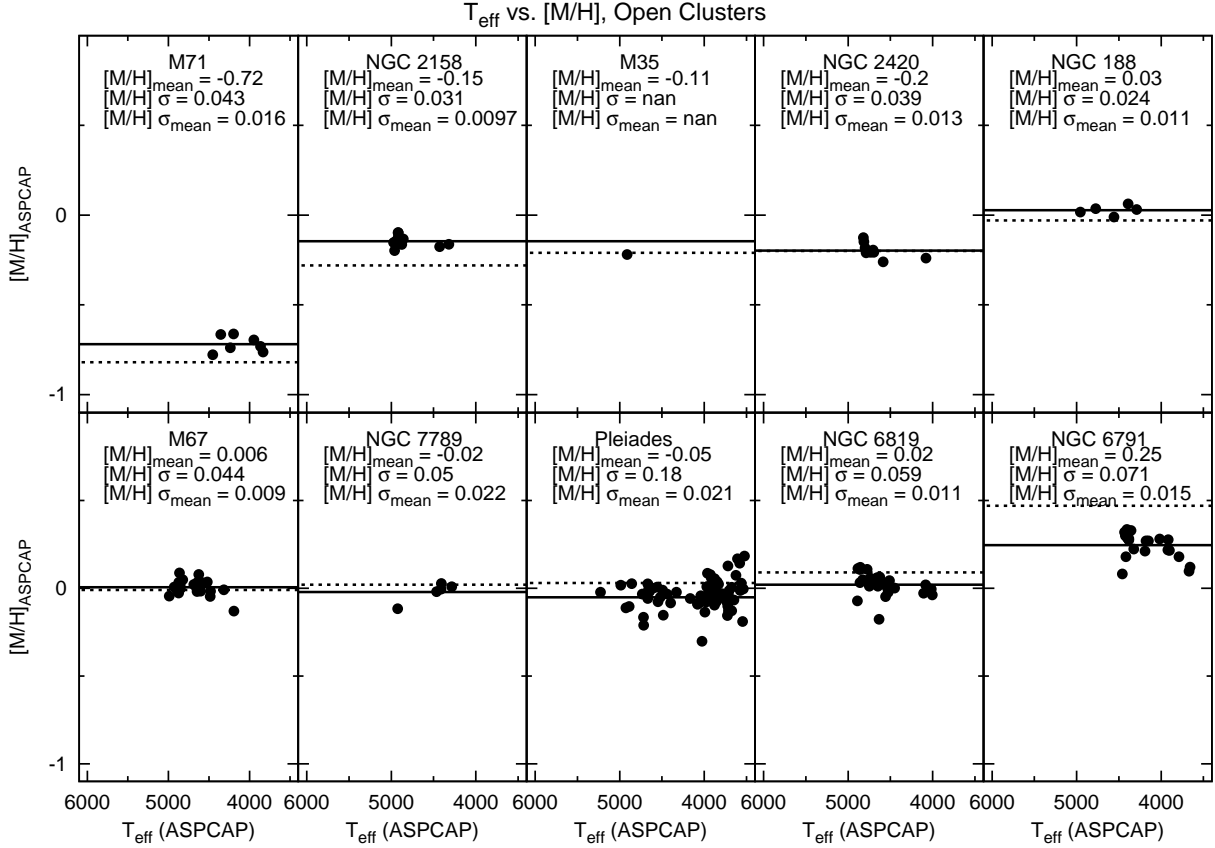}
\caption{Same as Figure 2, but for open clusters.  ASPCAP metallicities show good
agreement (to within less than 0.1~dex) with the literature
between [M/H]=$-$0.5 and [M/H]=$+$0.1, however at the very metal-rich
end they are underestimated by about 0.2$-$0.3~dex.  See Section
3.2 for discussion.
}
\label{fig:metopen}
\end{figure}

The ASPCAP metallicities were compared to literature values cluster
by cluster. Figure \ref{fig:metglob} (globular clusters) and 
\ref{fig:metopen} (open clusters) show the metallicity as a function of
effective temperature for all the clusters examined in this paper.
The average cluster metallicities adopted from the literature are
listed in Table 1. Table 4 lists the average cluster ASPCAP
metallicities and the standard deviation for each cluster. 

\begin{figure}[!ht]
\includegraphics[width=4.5in,angle=270]{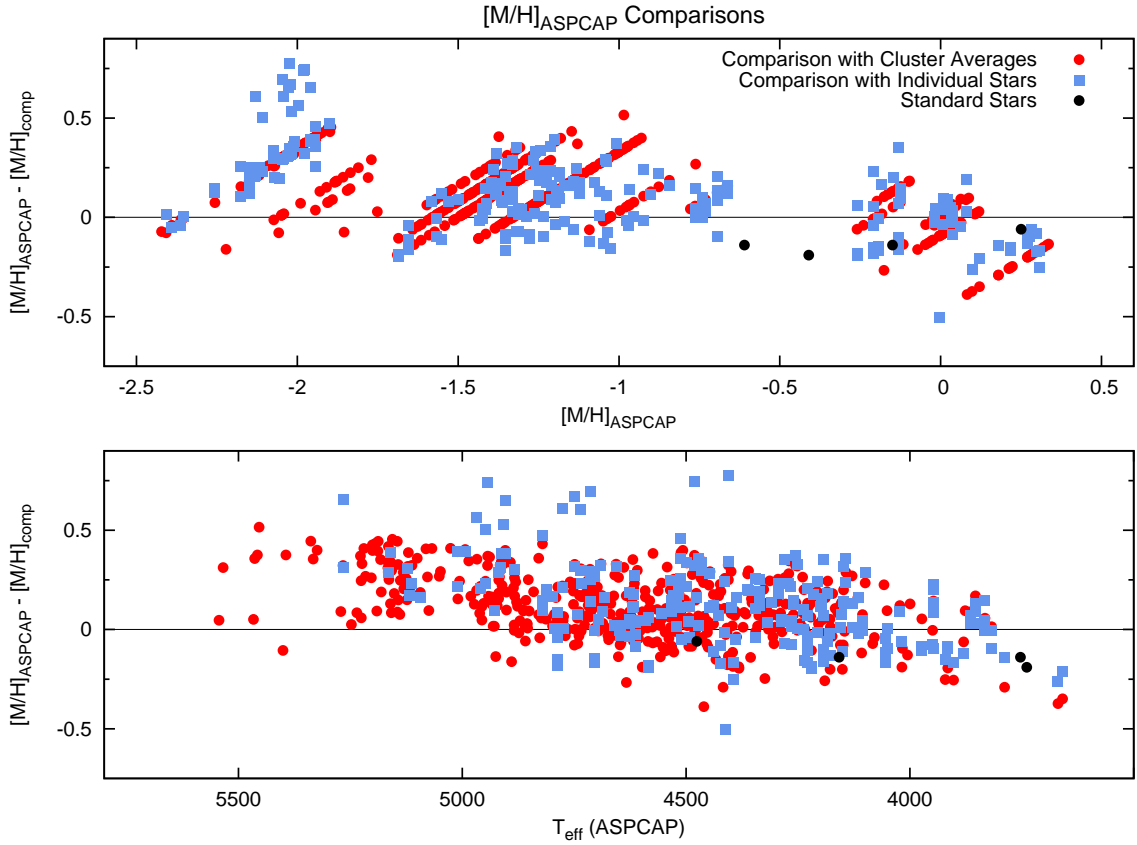}
\caption{ASPCAP [M/H] is compared to individual 
star by star (blue squares) metallicities for cluster members, and standard stars (black). The
differences from the cluster mean values are plotted with red dots
as a function of ASPCAP [M/H] (upper panel) and T$_{\rm eff}$ (lower panel). 
 Only giant stars with $\log~g$ $<$ 3.5 are plotted in both panels. }
\label{fig:metstar}
\end{figure}

\begin{figure}[!ht]
\includegraphics[width=4.5in,angle=270]{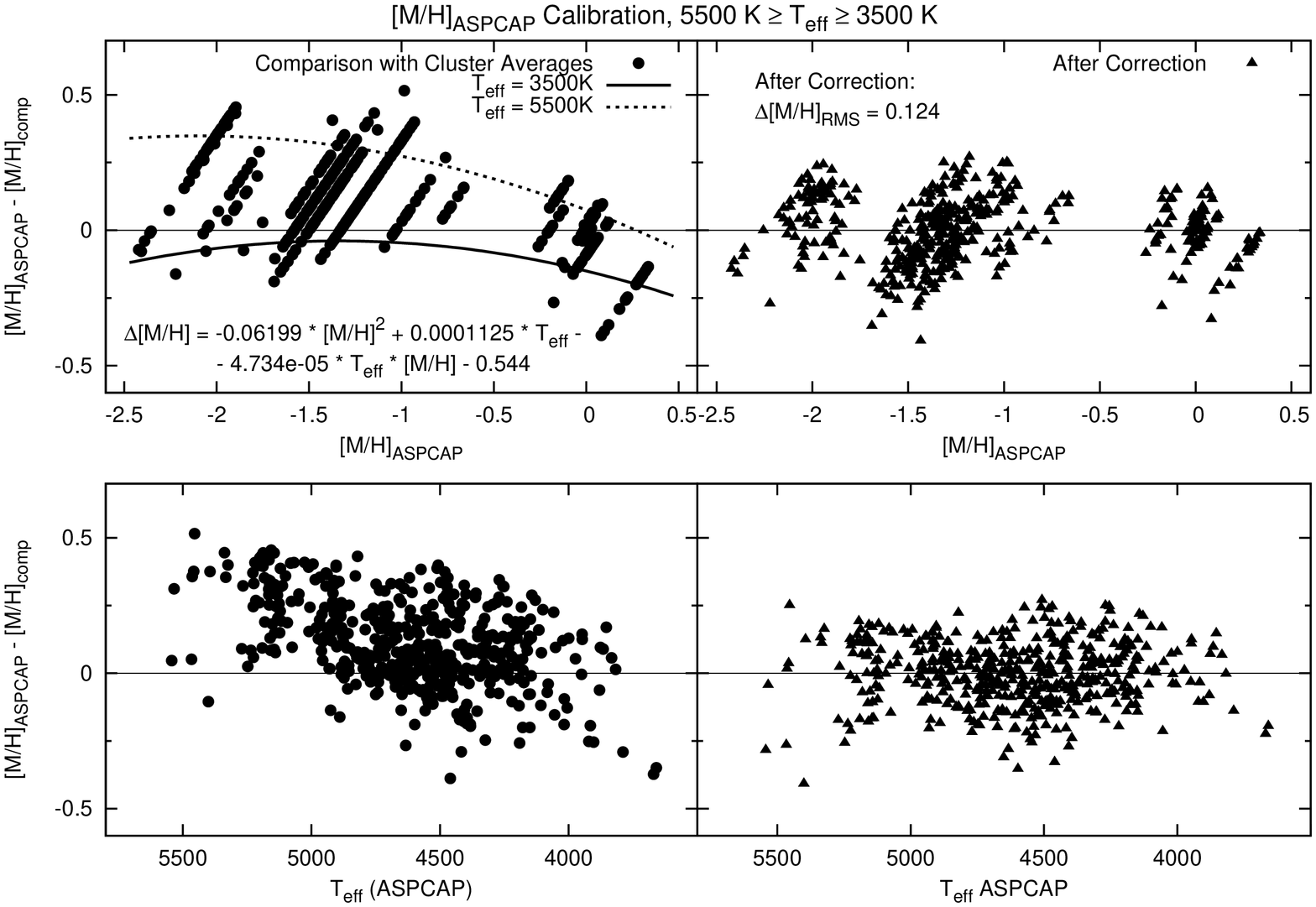}
\caption{
Left panels show the differences of raw ASPCAP metallicities and cluster 
averages plotted as a function of [M/H] (top) and 
T$_{\rm eff}$ (bottom). The solid line show the fitted relation (\ref{eqn:metallicity}) 
for 5500~K, and dotted lines show the same for 3500~K. Right panels show 
metallicities after calibration from \ref{eqn:metallicity}
is applied. See Section 3.2 for discussion. Only giant stars with $\log~g$ $<$ 3.5 are plotted in all panels. 
}
\label{fig:metcorr}
\end{figure}

Near solar-metallicity, the raw metallicities derived from the
APOGEE spectra are very close to the metallicities from the literature
(Figure \ref{fig:metopen}). Open clusters with $-0.5 < $[M/H]$ < 0.1$ agree within
0.1~dex with the literature averages. In globular clusters below
[M/H]$ = -0.5$, the differences increase as [M/H] decreases, from
systematic offsets of 0.1~dex for M71 to 0.35 for M92 (Figure \ref{fig:metglob}).
The reason that ASPCAP overestimates [M/H] at low metallicities may
be related to the decreasing number of metal lines, and increasing
number of $\alpha-$element lines (mostly OH), which leads to strong
correlations between the two quantities (see Section 3.5).  The
metallicity also shows indication of  linear trends as a function of T$_{\rm
eff}$ for almost all globular clusters. This is mostly visible in M92, M2, and
M13 (Figure \ref{fig:metglob}).  This dependence does not exist in the metal-rich
open clusters. Stars above 5000~K show a significant systematic
difference in T$_{\rm eff}$ (up to 200$-$500~K) compared to
photometric temperatures, and this contributes to systematically different
metallicities. 
This large temperature offset may combine with the error
coming from the lack of Fe lines at low metallicities, and the two
sources of errors together may lead to the overestimates of the metallicity and
small linear trends with temperature.

Figure \ref{fig:metstar} shows the differences between raw ASPCAP metallicities 
and the literature values using both the literature cluster averages (red) as well
as the measurements of individual cluster stars from the literature (blue). As
expected, the trends are similar, given that the cluster averages are determined from
the individual star measurements. Because the latter include star-by-star uncertainties,
we chose to use the cluster averages as a basis to derive an empirical calibration to 
bring the raw ASPCAP metallicities onto the literature scale.

\begin{figure}[!ht]
\includegraphics[width=4.5in,angle=270]{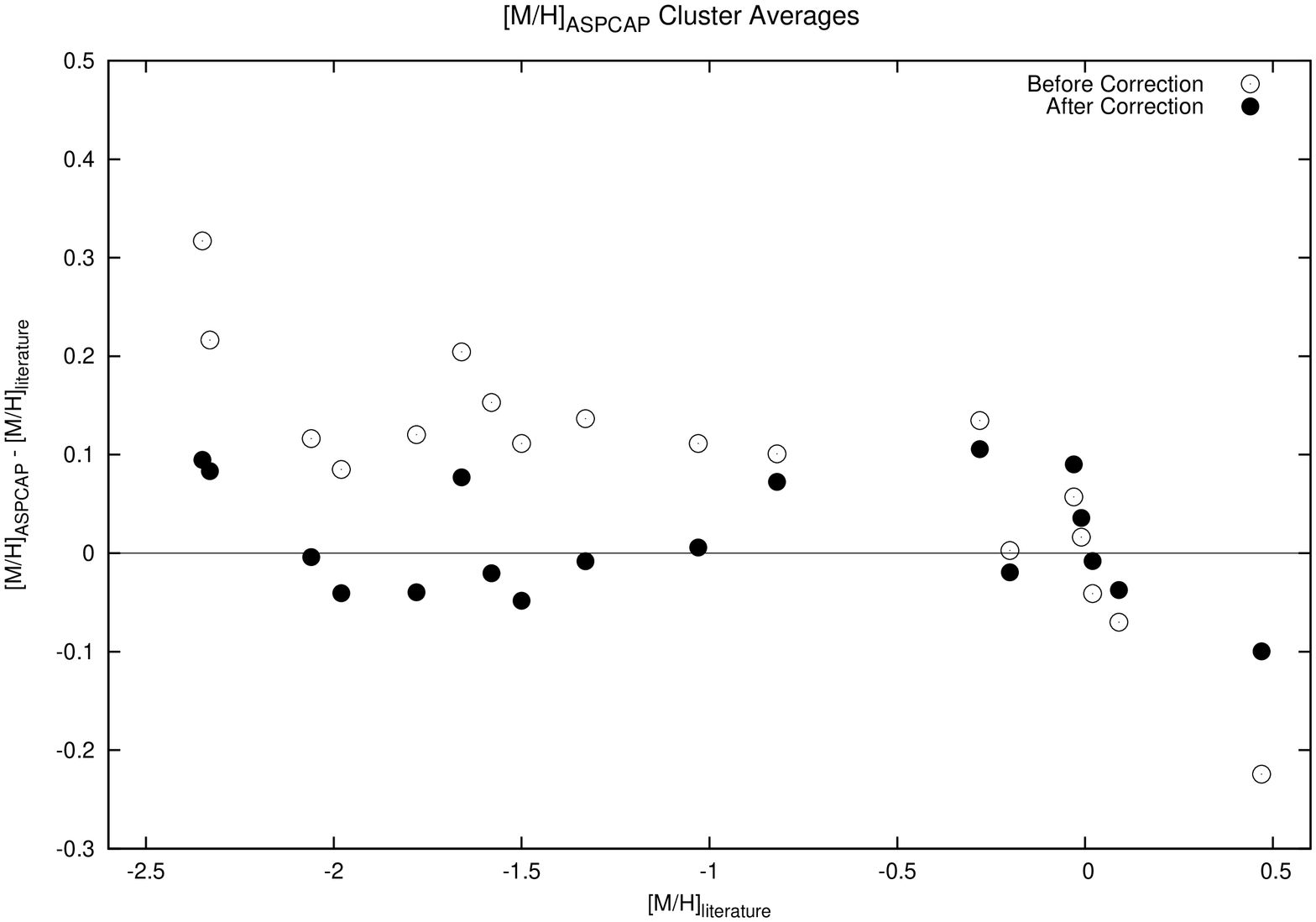}
\caption{ASPCAP cluster average metallicities before (open circles)
and after (filled circles) the correction. The corrected values lie
within $\pm$~0.1~dex from average literature values. The
latter are listed in Table 1, while ASPCAP values are presented in Table 4.
}
\label{fig:metclus}
\end{figure}

The ASPCAP metallicities for the standard stars (black points in Figure \ref{fig:metstar})
are too low compared with the analysis by \citet{smith01} by an average of 
0.13 $\pm 0.03$~dex as a result of the average $-$93~K difference in temperature.
According to \citet{gray01}, one might expect the abundances derived
from neutral lines (like in the case of the standard stars) to have
a dependence with temperature of $-1.3\times 10^{-3}$~dex~K$^{-1}$ when
the excitation potential is about 0 eV and $-0.4\times 10^{-3}$~dex~K$^{-1}$
when it is about 5 eV. In our case a 93~K shift corresponds to
0.12~dex, and 0.04~dex respectively. 
%Thus, we conclude that the
%differences in metallicity between ASPCAP and the manual analysis
%of FTS spectra by \citet{smith01} are due to the presence of a
%systematic effect as a result of different T$_{\rm eff}$ scales.

Within a given cluster, we found that the deviation of the ASPCAP
metallicity from the literature value was also a function of effective
temperature. The sensitivity to effective temperature is demostrated
in the bottom panel of Figure \ref{fig:metstar}. To account for both the
metallicity and temperature dependence of the systematic errors in
metallicity, we derived an empirical correction using both quantities
as independent variables:

\begin{equation}
[M/H]_{corrected} = [M/H]_A + 0.06199 [M/H]_A^2 - 1.125 \cdot 10^{-4} T_A + 
4.734 \cdot 10^{-5} T_A [M/H]_A + 0.544
\label{eqn:metallicity}
\end{equation}

where $[M/H]_A$ is the raw ASPCAP metallicity and $T_A$ is the raw ASPCAP effective 
temperature.
The calibration is valid between $-$2.5 and 0.5~dex in metallicity, and 
from 3500~K to 5500~K in temperature for all stars with $\log~g$ $<$ 3.5.

Figure \ref{fig:metcorr} shows the metallicities after this
correction is applied, demonstrating that this relation does a good
job of bringing the raw ASPCAP values into agreement with the
literature values. Because of the temperature term, the overall
scatter of the corrected differences is smaller as well, although
some trends with temperature at low metallicities are still present.  
The scatter of the differences after the correction
is 0.12~dex across the full metallicity range, which is only slightly
larger than the scatter of the original values in each of the
individual clusters (see Section 3.5).  Figure \ref{fig:metclus} shows the cluster
averages compared to the literature values before and after correction.
Cluster averages in the full metallicity range are within $\pm$~0.1~dex
the literature after calibration.

\subsection{Accuracy of Surface Gravity}

Spectral features are generally less sensitive to surface gravity
than to effective temperature or metallicity, so accurate
values are challenging to derive. However, surface gravities are a
critical ingredient for the estimation of  spectroscopic
parallaxes, therefore achieving the highest possible accuracy is
important. Simulations based on photon noise-added synthetic
spectra predict a floor of approximately 0.2~dex in $\log~g$
uncertainty for APOGEE spectra with S/N$\sim$100/pixel, although
this is a function of effective temperature and metallicity.

\subsubsection{Surface Gravity from Isochrones and Stellar Oscillations}

To check the accuracy of our derived surface gravity, several
types of independently derived gravities were used.

For cluster stars, theoretical isochrones can be used to provide
a surface gravity for a given effective temperature, if the
cluster age and distance are known. We derived gravities for
our sample using isochrones from the Padova group \citep{bertelli01, bertelli02},
adopting cluster parameters as given in Table 1; because 
the Padova isochrones use scaled-solar abundances, we adopted
isochrones with metallicities increased by 0.2 over the adopted [Fe/H]
for the metal-poor globular clusters to account for the alpha-enhancement
generally seen in these. We compared these results to those obtained using
isochrones from the Dartmouth group \citep{dotter01}, and found
that the derived surface gravities were within 0.1 dex, i.e. the differences are smaller than
our expected random errors. For effective temperatures, we
used the calibrated ASPCAP temperatures discussed above; as described
below, the calibrated temperatures provided more consistent results
than those obtained using the raw ASPCAP temperatures. We note, however, that surface
gravities derived in using isochrones have significant uncertainties: they
are correct only to the extent that model isochrones on the giant
branch are accurate, and even then, since the giant branches are steep,
a 100~K error in temperature leads to a 0.2$-$0.3~dex error in gravity 
depending slightly on metallicity. 

The second method involves a comparison of ASPCAP gravities for
stars in the Kepler field with those derived from asteroseismic
analysis. These are expected to be highly accurate, with uncertainties of 
0.01$-$0.03 dex. However, the Kepler stars do not include any metal-poor
objects and thus do not span the full range of parameters of
APOGEE stars.

Further verification of the surface gravities can be made by
comparing the ASPCAP results with those determined by
other spectroscopic studies for stars in which such data exist.
However, log~g values from the
literature generally have significant uncertainties (up to
0.2$-$0.4~dex), so we believe that this approach provides
poorer calibration than either isochrone or asteroseismic comparisons.

\subsubsection{Empirical Calibration of Surface Gravity}
\label{sect:logg}

\begin{figure}[!ht]
\includegraphics[width=4.5in,angle=270]{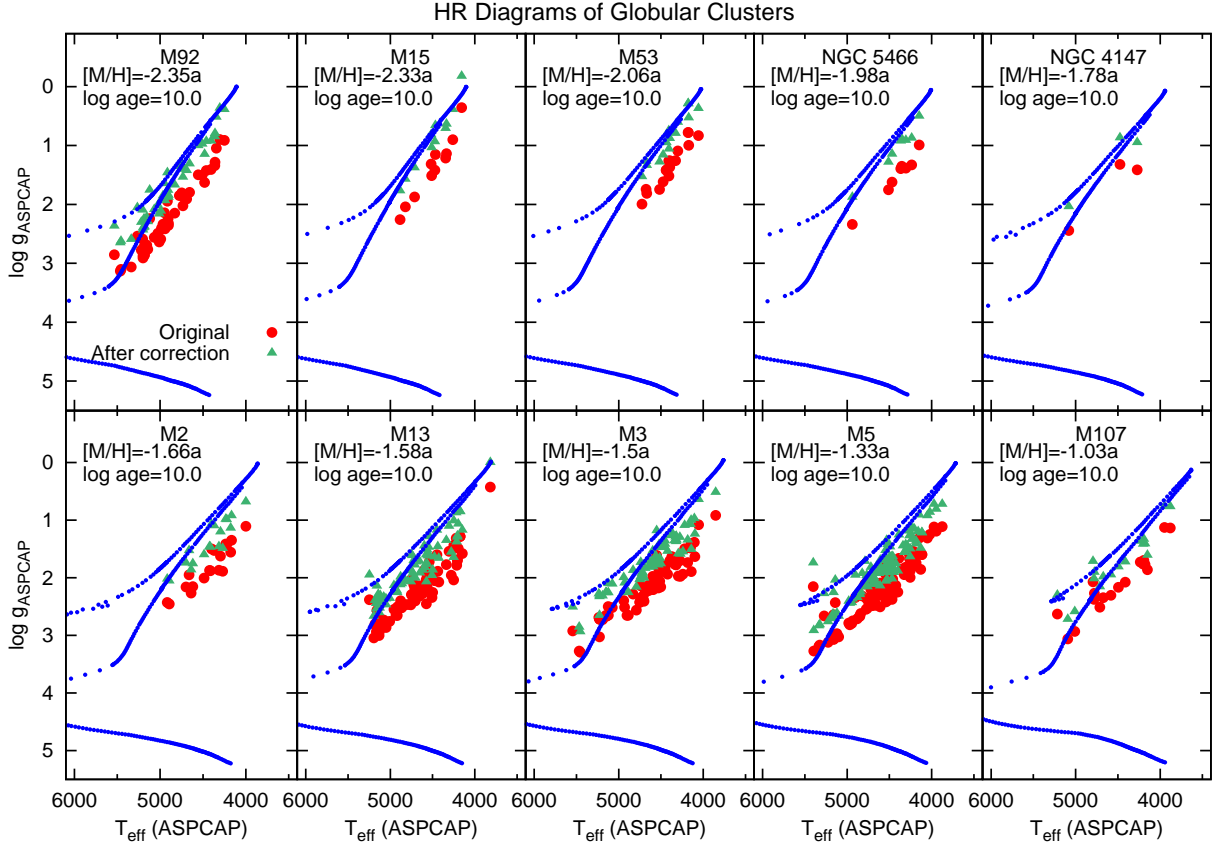}
\caption{T$_{\rm eff}-$ $\log~g$ diagrams of globular clusters.
Metallicities denoted by "a" in the labels were increased by 0.2~dex
to account for the increased $\alpha$ content, and these increased
values were used to generate the plotted isochrones.  Red dots show
the original ASPCAP $\log~g$ values as a function of ASPCAP T$_{\rm
eff}$. Isochrones and \textit{Kepler} stars were used to calibrate
ASPCAP gravities, and new values after the calibration described in
Section 3.3 are plotted with green triangles.  }
\label{fig:hrglob}
\end{figure}

\begin{figure}[!ht]
\includegraphics[width=4.5in,angle=270]{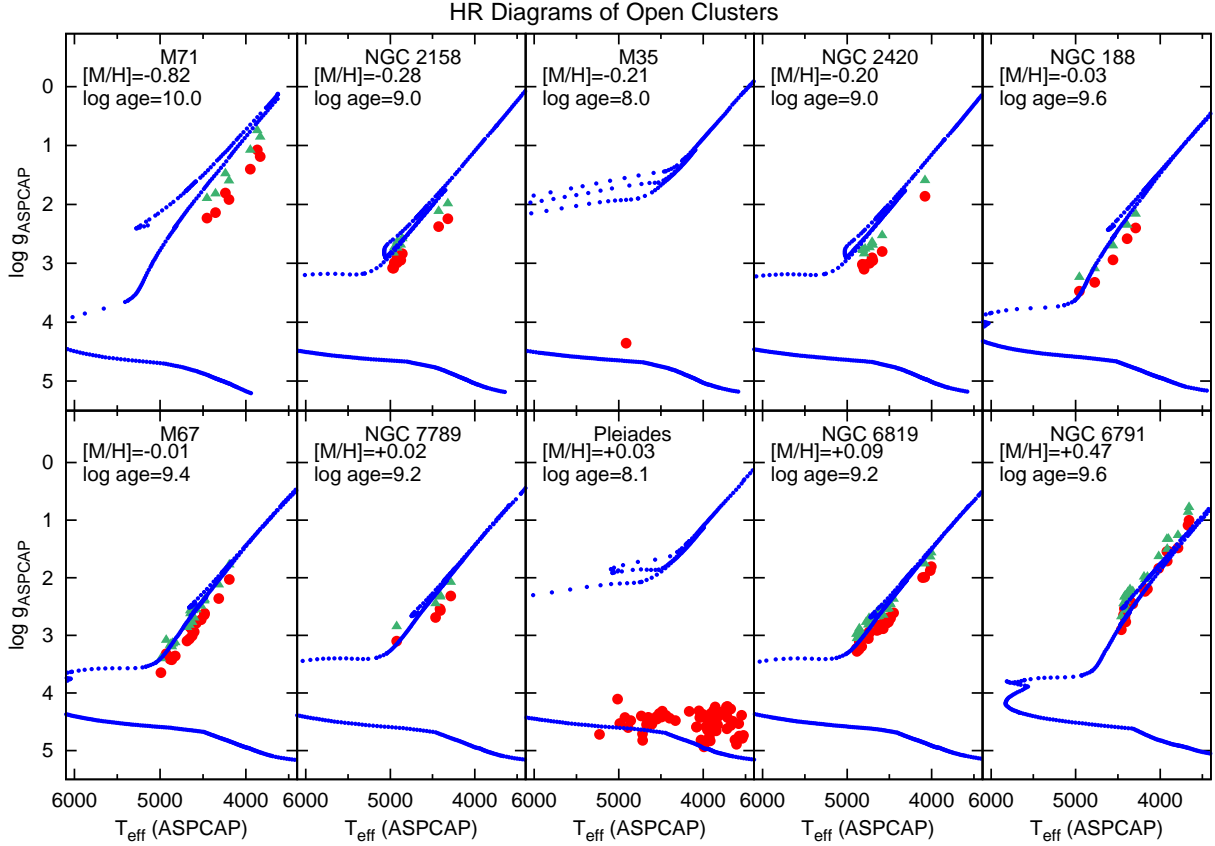}
\caption{T$_{\rm eff}-$ $\log~g$ diagrams of open clusters. For
explanation please see the caption of Figure \ref{fig:hrglob}.  }
\label{fig:hropen}
\end{figure}

In Figures \ref{fig:hrglob} and \ref{fig:hropen}, ASPCAP surface gravities and 
effective temperatures (red dots) are shown along
with theoretical isochrones (blue line) for a number of globular
and open clusters. Significant disagreement is apparent
for the metal-poor globular clusters.  We note that while the
temperature correction discussed above would move the ASPCAP
points into slightly better agreement with the isochrones, it
is not large enough to account for the bulk of the discrepancy:
ASPCAP surface gravities are inferred to be $\sim$ 0.5 dex too high for metal-poor
stars. For the more metal-rich open clusters (Figure \ref{fig:hropen})
the ASPCAP points fall closer to the isochrones, suggesting
offsets of 0$-$0.3 dex in surface gravity.

\begin{figure}[!ht]
\includegraphics[width=4in,angle=0]{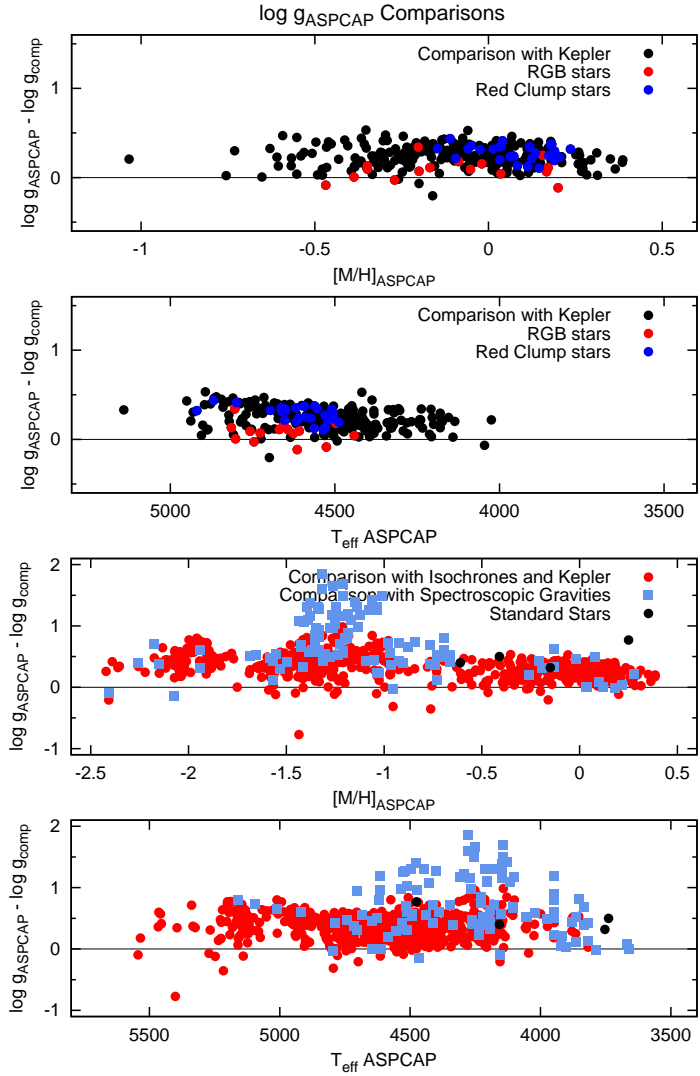}
\caption{\textit{Top two panels}: Comparison of ASPCAP log g with asteroseismic log g for
stars in the Kepler fields. Top panel shows differences as a function
of ASPCAP metallicity, while bottom panel shows differences as function
of temperature. Red points are asteroseismically confirmed giants, while
blue points are asteroseismically confirmed red clump stars, while the evolutionary status of stars denoted by black
dots are not yet confirmed.
\textit{Bottom two panels}: ASPCAP gravity differences compared to isochrones and
\textit{Kepler} stars (red dots), spectroscopic $\log~g$ (blue
squares), and standard stars (black dots) are plotted as a function
of [M/H] (upper panel) and T$_{\rm eff}$ (lower panel). In case of the 
isochrones, corrected temperature were used to determine the 
comparison value of surface gravity. Only giant stars with $\log~g$ $<$ 3.5 are plotted in all panels. 
}
\label{fig:gravdiff}
\end{figure}

Comparison of ASPCAP and asteroseismic gravities for Kepler stars
is shown in the upper panels of Figure \ref{fig:gravdiff}, where the difference
between the two sets of gravities is plotted against metallicity 
and effective temperature. This suggests that
ASPCAP gravities are generally too high by a few tenths of a dex.
Since, as noted above, the Kepler sample does not include more
metal-poor stars, it cannot be used to confirm the larger discrepancies
suggested by the isochrone comparison for metal-poor clusters. 
Figure \ref{fig:gravdiff} suggests that there is substantial scatter
in the difference between ASPCAP and asteroseismic log g that
is not well correlated with either metallicity or temperature, although
there is certainly a trend with temperature for the bulk of the sample.
Interestingly, the deviations appear to depend on evolutionary
state of the star: red points denote stars that are asteroseismically
determined to be hydrogen burning RGB stars, while blue points are
helium burning red clump stars. We do not yet have a good explanation 
for this.

Generally, the inferred offsets in surface gravity are comparable
using the isochrone gravities and the asteroseismic gravities, but
there are some small inconsistencies. For the most metal-rich stars,
the isochrones suggest that the ASPCAP surface gravities are nearly
correct, while the asteroseismic analysis suggests that the ASPCAP
gravities are still overestimated. In addition, the isochrone
comparison does not suggest trends with effective temperature that
appear for a large fraction of the Kepler sample.  We note that small changes
in the ASPCAP temperature scale or in adopted cluster parameters
are unable to resolve these discrepancies.  It is possible that they
are related with the issues with evolutionary state.

\begin{figure}[!ht]
\includegraphics[width=4.5in,angle=270]{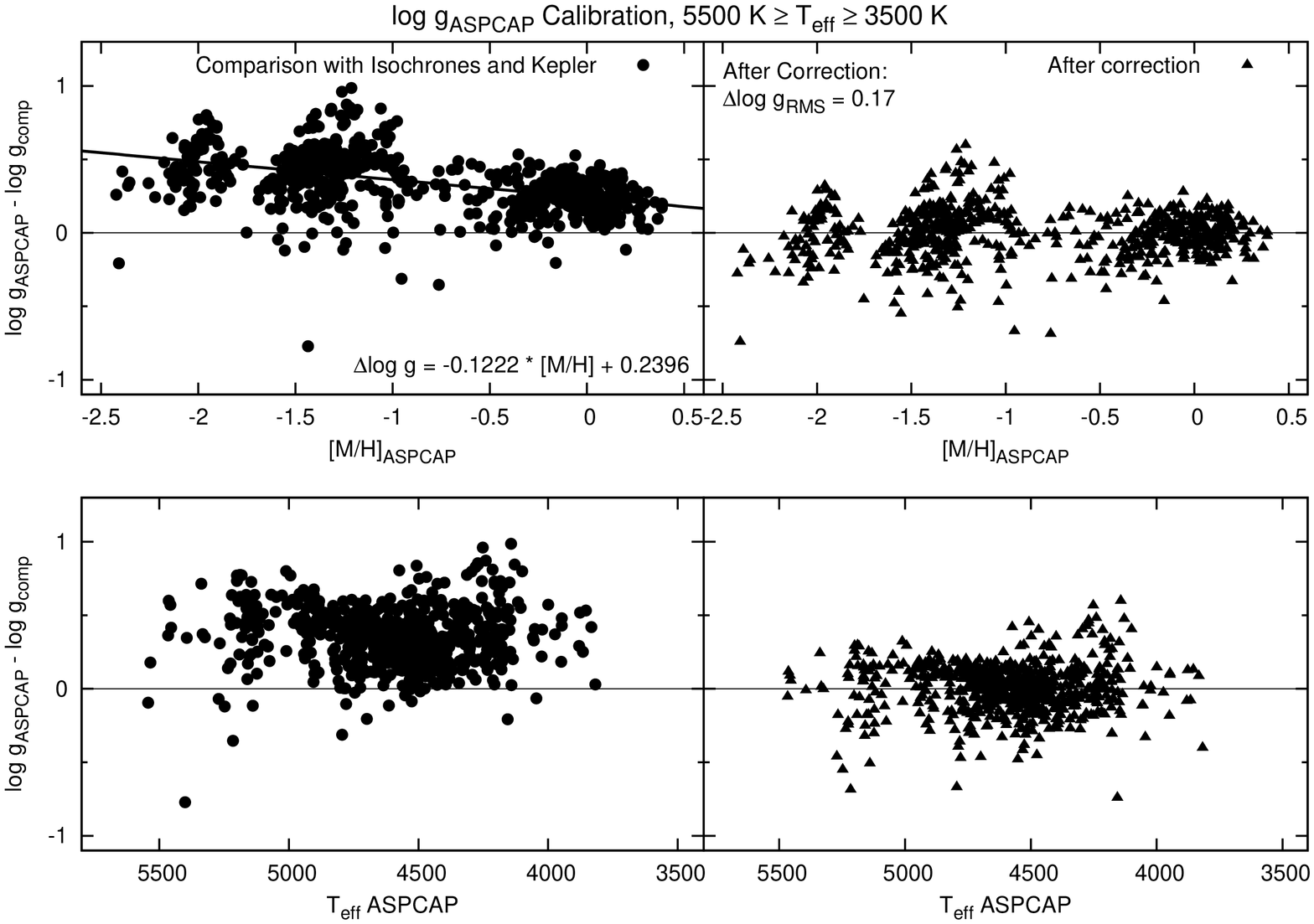}
\caption{Comparison of ASPCAP log~g with the combined dataset of asteroseismic and isochrone surface gravities. 
The applied empirical correction is shown on the left panels. Results after the
empirical calibration are shown on the right panels. Values after the correction
results in a 0.17~dex scatter covering the full metallicity and
temperature range. See Section 3.3 for discussion. Only giant stars with $\log~g$ $<$ 3.5 are plotted in both panels. 
}
\label{fig:gravcorr}
\end{figure}

To derive a calibration relation for surface gravity, we adopt
a combined sample of Kepler stars with asteroseismic surface
gravities and metal-poor cluster stars with isochrone surface gravities;
specifically, we use all of the Kepler
stars with $[M/H]>-1$ and all of the clusters with $[M/H]<-0.5$ (so
there is an overlap region where both clusters and Kepler stars
are used). We used corrected temperatures to determine gravities 
from isochrones, 
because adoption of original ASPCAP values increases  the disagreement 
with seismic gravities. 

The lower two panels of Figure \ref{fig:gravdiff} show 
how the $\log~g$ differences depend on metallicity
and T$_{\rm eff}$ for this sample, and also includes a comparison of ASPCAP
gravities with independently derived optical spectroscopic gravities from
the literature (blue points) and from the fundamental (based on parallaxes) values 
for the standard
stars \citep[black dots in the two bottom panels of Figure 9.,][]{smith01}. 
These generally agree with the differences
inferred from isochrone and asteroseismic analysis, although there is
significant scatter. The standard stars show
no significant trends with metallicity or temperature, and an
average difference of $0.48 \pm 0.08$ dex, but we have to note that the gravity of $\mu$ Leo derived 
by \citet{smith01} is 0.77~dex lower than that determined by ASPCAP;
this is a much larger deviation than suggested by any other data
at this metallicity.

We adopt a gravity correction that is a function of metallicity only,
as including any other independent variables is unable to improve
the agreement.
Fitting the observed differences yielded the empirical calibration
relation:

\begin{equation}
\log~g_{corrected} = \log~g_A + 0.1222 [M/H]_A - 0.2396
\label{eqn:logg}
\end{equation}
where [M/H]$_A$ is the raw ASPCAP metallicity and $\log~g_A$ is
the raw ASPCAP surface gravity.

Figure \ref{fig:gravcorr} presents the values before (left panels) and 
after (right panels) the calibration. For all metallicities and for 
T$_{\rm eff}$ between 3800 and 5500~K, this empirical correction reduces most of
the systematic differences to around zero.
The RMS scatter of the gravity differences compared with isochrones
and \textit{Kepler} gravities is reduced to 0.17~dex.  
We attempted to derive a calibration
relation using other functional forms, but we found that the above
solution gave the smallest standard deviation. We recommend that this 
relation be used, but with application limited to $3800 < T_A <
5500$~K, $-2.5 < $[M/H]$ < 0.5$~dex, and $\log~g < 3.5$.

\begin{figure}[!ht] 
\includegraphics[width=4.5in,angle=270]{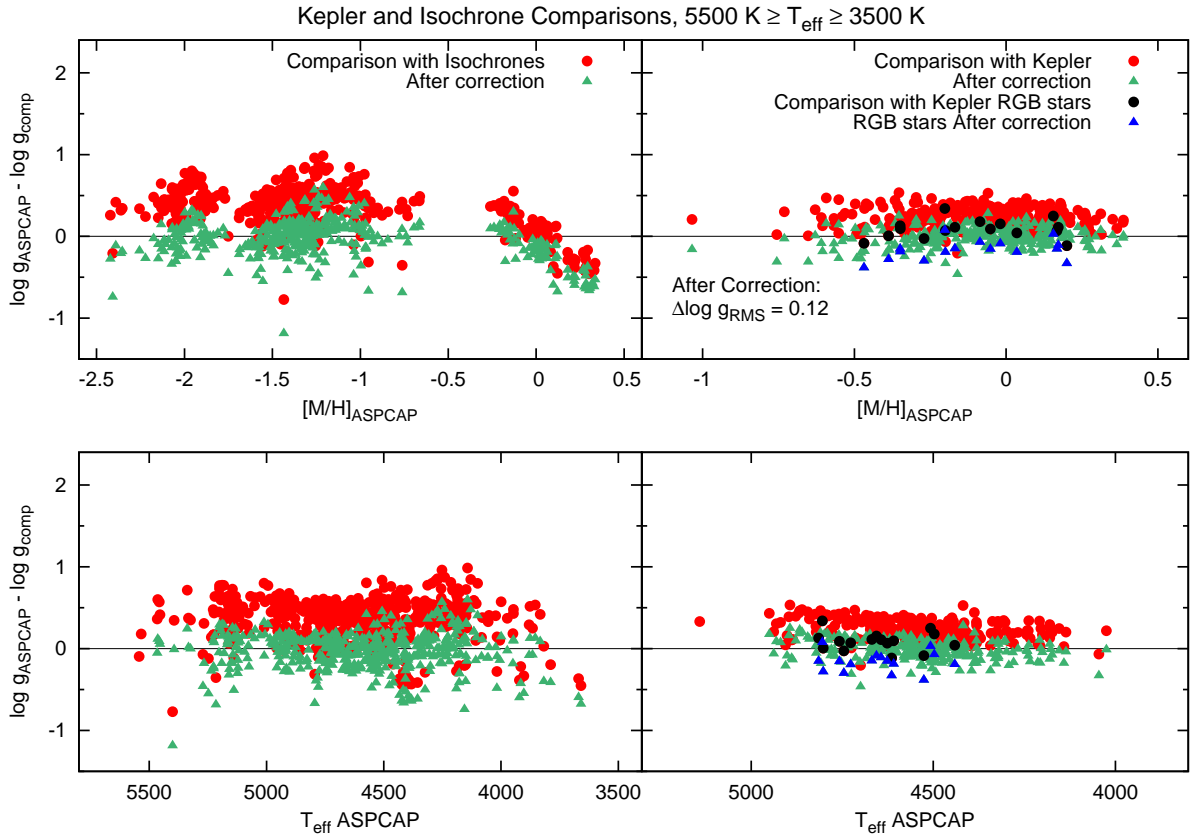}
\caption{ The calibration
from Figure \ref{fig:gravcorr} applied to only isochrone differences (left panels),
and to the \textit{Kepler} sample (right panels). Some discrepancy remain at solar 
metallicities, where the calibration results in poor agreement with isochrone surface 
gravities in open clusters. Confirmed
\textit{Kepler} RGB stars \citep{mosser01} are also plotted before
(black dots) and after calibration (blue triangles). The evolutionary state 
classifications were done by \citet{stello01}. Only giant stars with $\log~g$ $<$ 3.5 are plotted in all panels.  } 
\label{fig:gravcorr2}
\end{figure}

As a sanity check, in Figure \ref{fig:gravcorr2} we compare the application of this calibration 
separately to the isochrone gravities (left panels) and to 
asteroseismic gravities (right) panels.
The RMS scatter is only 0.12~dex after the correction for
the \textit{Kepler} stars, which is a significant improvement.
The error in surface gravity shows a small dependence
on temperature (Figure \ref{fig:gravcorr2} lower right panel); this
trend remains after application of the calibration relation.
The discrepancy between corrections based on isochrone and asteroseismic 
gravity at high metallicity is evident in the left panel, where the
corrected values show some disagreement with the isochrone gravities,
in particular at the highest metallicities.

Figures \ref{fig:hrglob} and \ref{fig:hropen} show the corrected
points in the HR diagrams as green triangles.
The corrected values are closer to the isochrones, but
some discrepancies still remain, especially at low
metallicities around [M/H]=$-$1 to $-$1.5 in M5, M3, and M13 (Figure
\ref{fig:hrglob}, lower panels), and at high metallicities above +0.2~dex in
NGC~6791 (Figure \ref{fig:hropen} lower right panel).

\subsection{Accuracy of $\alpha-$elements, Carbon and Nitrogen Abundances}

\begin{figure}[!ht]
\includegraphics[width=4.5in,angle=270]{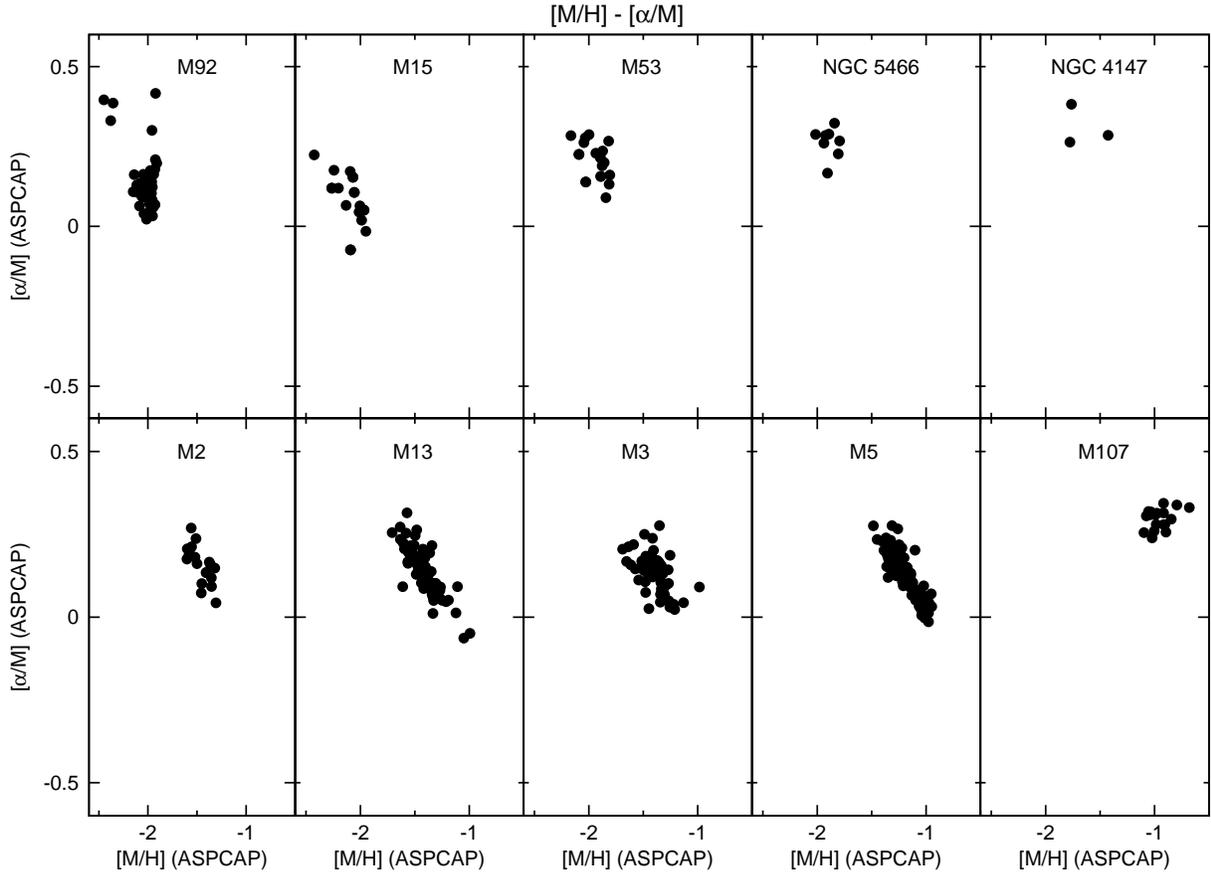}
\caption{ASPCAP [$\alpha$/M] as a function of metallicity for globular clusters. 
Strong correlations between the two parameters is visible in all globular clusters.
}
\label{fig:alphaglob}
\end{figure}

\begin{figure}[!ht]
\includegraphics[width=4.5in,angle=270]{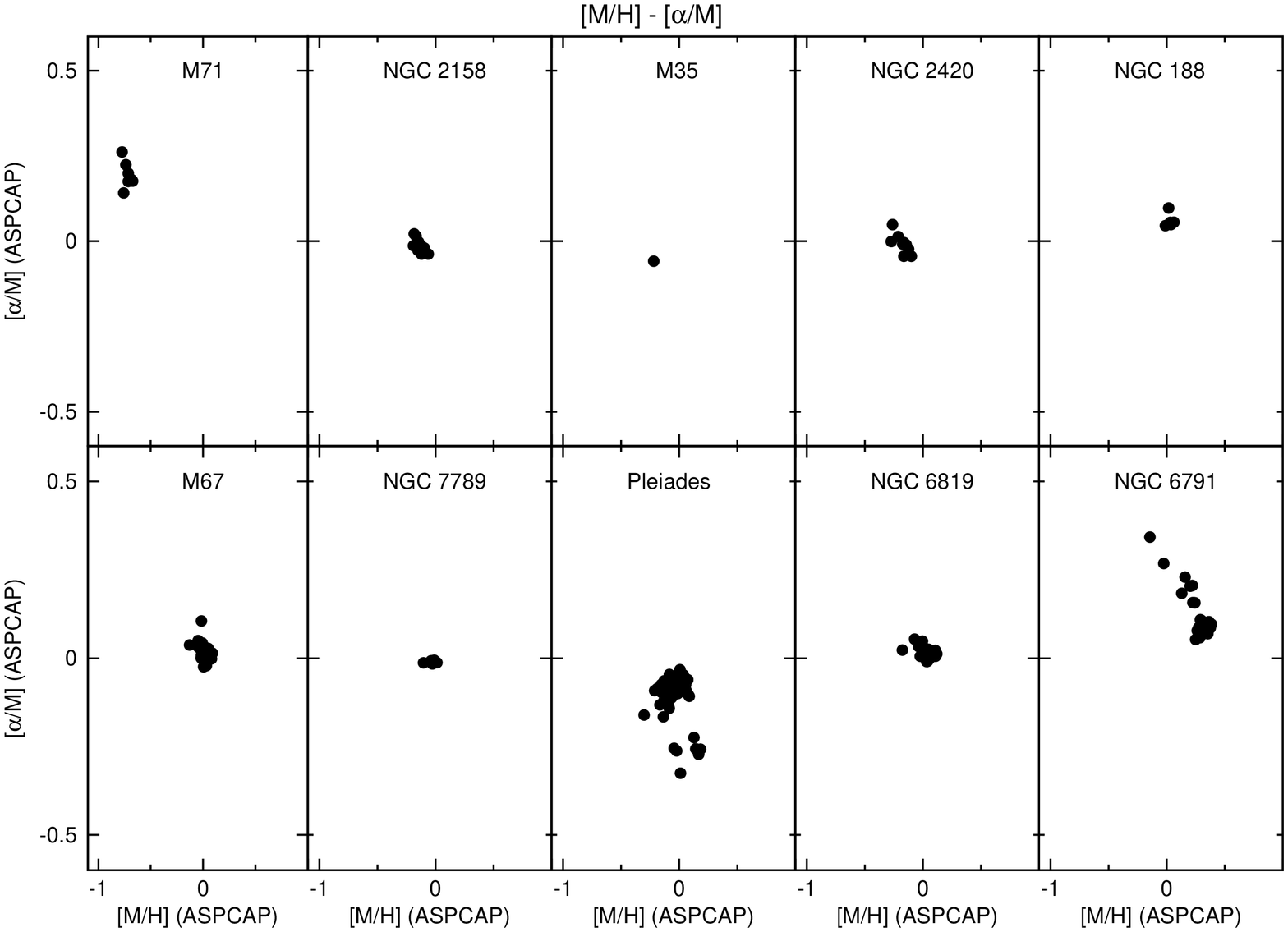}
\caption{ASPCAP [$\alpha$/M] as a function of metallicity for open clusters. The
[$\alpha$/M] abundances close to solar metallicities show low scatter
and have no visible dependence on metallicity or temperature. 
}
\label{fig:alphaopen}
\end{figure}

Besides effective temperature, metallicity and gravity, three other
parameters are also published in DR10: $\alpha-$elements, carbon,
and nitrogen abundances. These six parameters are are simultaneous
derived by ASPCAP. 
Figures \ref{fig:alphaglob} and \ref{fig:alphaopen} show
[$\alpha$/M] as a function
of metallicity for the globular and open clusters.
In the metallicity region important for APOGEE ([M/H]
$> -$0.5 and $< +$0.1), the derived $\alpha$ abundances show no
trend in individual stars within a cluster with either metallicity,
or temperature.  However, strong dependencies are present as a
function of [M/H], 
as well as T$_{\rm eff}$ (not shown), both below [M/H] = $-$0.5
(shown in M13, M5 for example) and above [M/H] = $+$0.1 (shown in NGC~6791). 
The trend in NGC~6791 is not supported by other observations, as  
abundances of $\alpha-$elements in NGC~6791 from the literature
\citep{origlia01, carraro01} are between [$\alpha$/M]=$-$0.1 and
+0.1~dex, which indicates no $\alpha$ enrichment in this cluster.

\begin{figure}[!ht]
\includegraphics[width=4.5in,angle=270]{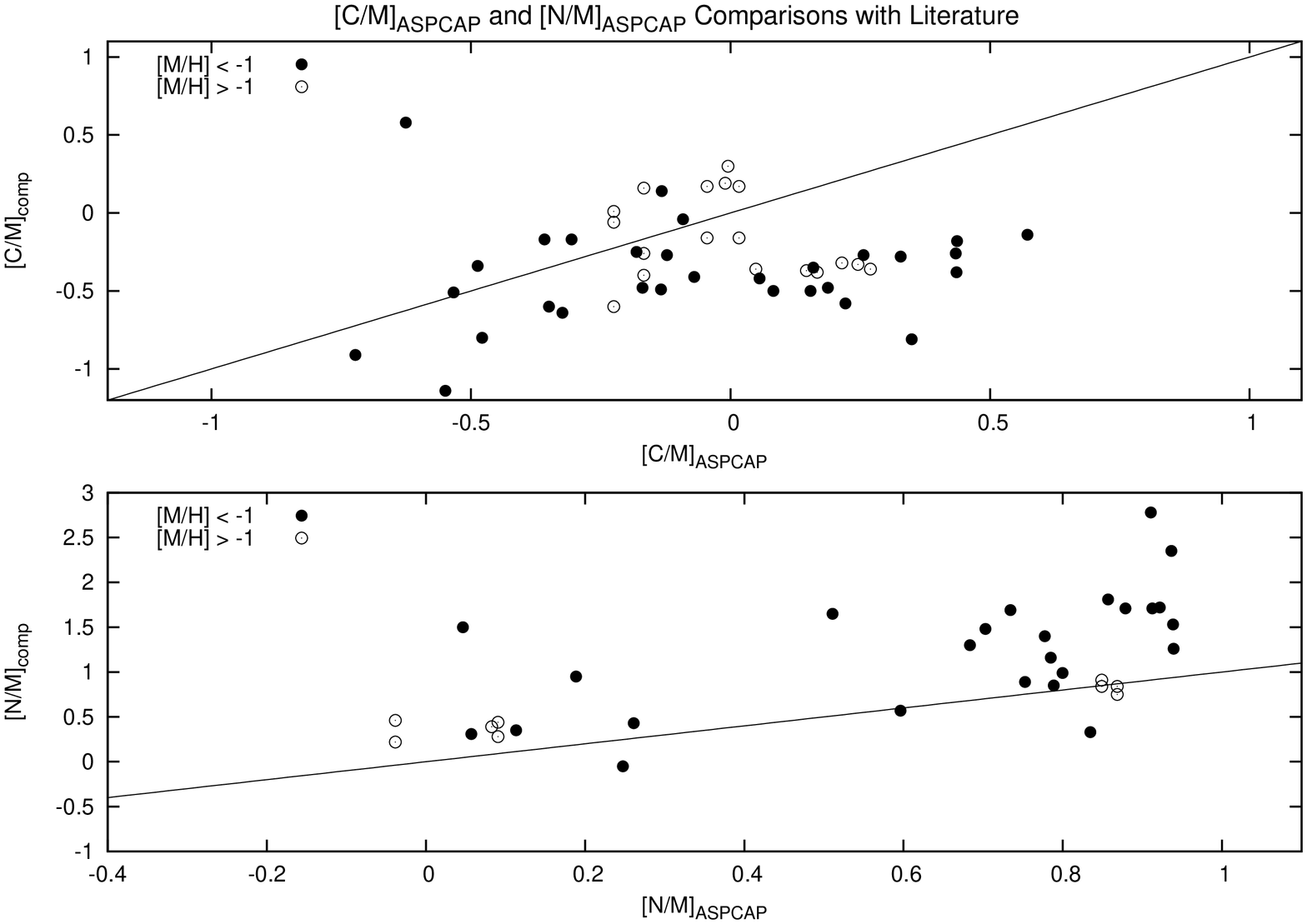}
\caption{C and N abundances compared to literature values. Metal
poor clusters below [M/H] $< -1$ are marked with solid circles,
while metal-rich clusters above -1 are denoted by open circles. See
Section 3.4 for discussion.  }
\label{fig:cn}
\end{figure}

Some of the above ambiguities may be due to the paucity and the weakness
of metal lines in the spectra of metal-poor stars.  Since few
iron lines are visible while OH lines remain strong in metal-poor
stars, the strength of the OH lines can be similarly matched by increasing
[M/H] or $[\alpha$/H].  This situation is illustrated in
Figure \ref{fig:alphaglob}, where an anti-correlation between [$\alpha/$Fe] and [M/H]
is apparent for example in the clusters M13 and M5. This effect is 
likely caused by correlated errors in metallicity and $\alpha$ 
abundances for NGC~6791.

The carbon and nitrogen abundance comparisons with literature values  
are shown in Figure \ref{fig:cn};
Table 1 lists the references used for the literature values.
Carbon abundances are significantly overestimated for globular
clusters (up to 1~dex), while for open clusters the agreement is
better between [C/M] = $-$0.5 and 0. Nitrogen abundances are generally
underestimated compared to other studies by about 0.2$-$0.4~dex at
low nitrogen content, and up to 1~dex for stars with high nitrogen
content. The overestimate of [N/M] at high [C/M] values 
may be a consequence of deriving nitrogen from CN: if ASPCAP underestimates
[C/M] (as shown in the upper panel for [C/M] $>$ 0), [N/M] will thus
be overestimated. We note that these comparisons are affected
by the relative paucity of [C/Fe] and [N/Fe] data for
cluster stars in the literature and by their intrinsic
uncertainties, particularly in the case of nitrogen.  Therefore,
it is difficult to decide how much of the large scatter and zero
point differences seen in Figure \ref{fig:cn} is contributed by the APOGEE
data, and how much by the literature data.  Because of this
uncertainty, we suggest that the current carbon and nitrogen
abundances derived by ASPCAP be used with extreme caution.
While the uncertainties in the C and N abundances may contribute
to systematic effects in other derived parameters, in ways that are still
not entirely understood, these possible systematics have been 
approximately removed by the calibration
process presented in this paper. Thus, we expect the calibrated values
for effective temperature, surface gravities, metallicities, and
$\alpha$-element abundances to be free of any important systematic
effects.

We believe that by defining certain spectral windows in the H-band
for all these elements, along with other ASPCAP improvements, is likely
to significantly improve the quality of the abundance determinations 
for these elements.

\subsection{Precision of Effective Temperature, Metallicity, and $\alpha-$elements}

\begin{figure}[!ht]
\includegraphics[width=4.5in,angle=270]{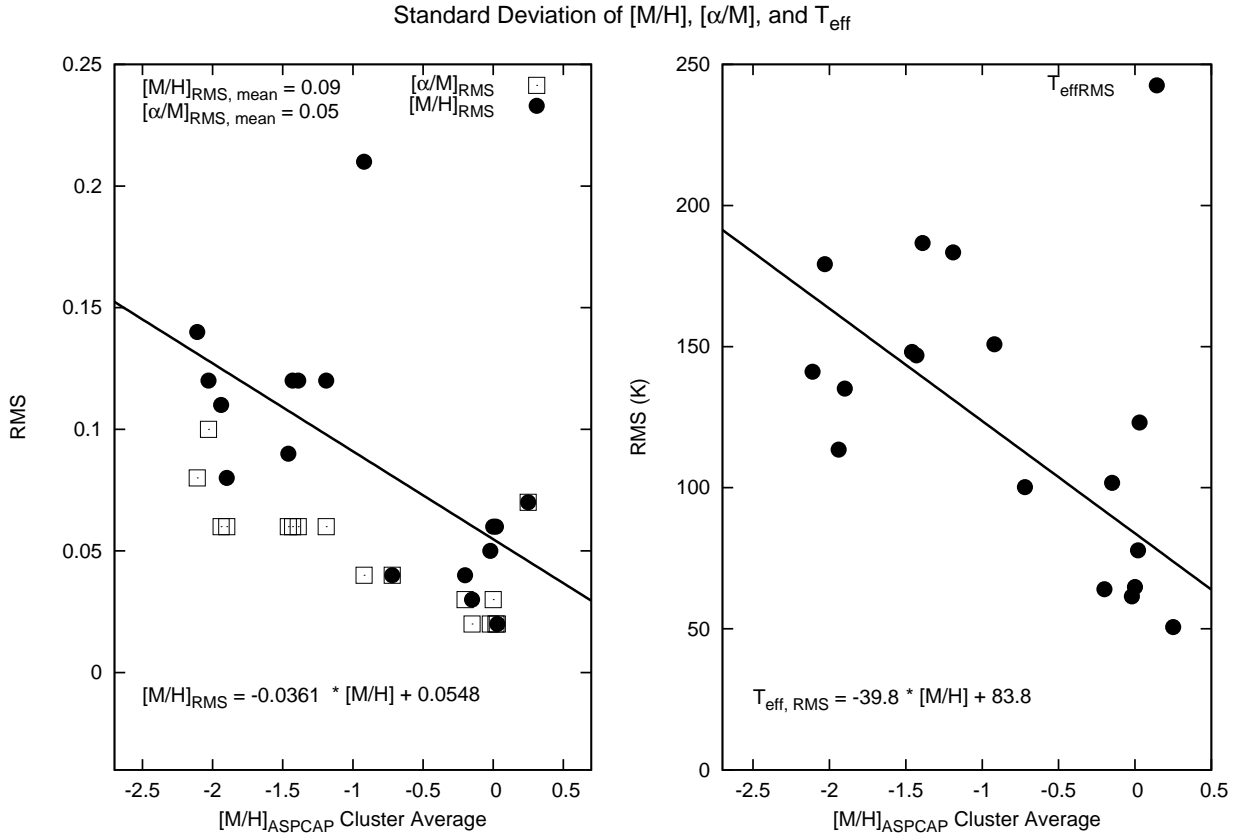}
\caption{The RMS of [M/H], [$\alpha$/M] and T$_{\rm eff}$ for each cluster 
(based on stars with $\log~g$ $<$ 3.5) as a function of
uncorrected cluster average metallicity from ASPCAP. A linear empirical calibration was used in case of 
effective temperature and metallicity. 
See Section 3.5 for discussion.
}
\label{fig:err}
\end{figure}

ASPCAP provides formal internal
error estimates for stellar parameters that are presently unrealistically small.
The scatter within individual clusters offers  
an external estimate of uncertainties, which we adopt 
for T$_{\rm eff}$ and metallicity.  The precision of the 
effective temperatures was calculated using the differences compared
to the $J-K_s$ calibrations. Any systematic shifts were neglected by
fitting a constant value to the differences. The error in temperature
for each cluster was then calculated by determining the RMS scatter around 
this fit. The calculated uncertainties are shown in the right panel of
Figure \ref{fig:err}, and listed in Table 4. 
The overall scatter in temperature is less than 200~K, but
it appears to be significantly better for metal-rich stars, where it is
generally smaller than 100~K. A linear fit was used to 
characterize how the precision depends on metallicity, and the 
following function was adopted to estimate the ASPCAP temperature errors:

\begin{equation}
T_{eff, RMS} = -39.8 \ [M/H]_A + 83.8 
\label{eqn:terr}
\end{equation}
where [M/H]$_A$ is the raw ASPCAP metallicity. The equation is valid for stars with $\log~g$ $<$ 3.5.

The RMS scatter of metallicity, shown by circles in the left panel of Figure \ref{fig:err},
was again determined by calculating the standard deviation around the 
cluster average values. The scatter in metallicity, in the range 0.08$-$0.14 dex for globular 
clusters, is higher for the most metal-poor clusters. For all the open clusters, with a nearly solar 
metallicity, the scatter is smaller, at about 0.03$-$0.07 dex. The average scatter 
using all clusters is 0.09~dex. There is one outlier cluster, M107, in which the scatter is 0.21~dex. This might
be the result of possible uncertainties in metallicity above 5000~K, due to increased temperature offsets from
photometric values. All other derived atmospheric parameters for M107 behave just like the other
globular clusters. The RMS for the ensemble of clusters clearly depends on the metallicity, 
thus again a linear equation was used to estimate the ASPCAP metallicity errors:

\begin{equation}
[M/H]_{RMS} = -0.0361 \ [M/H]_A + 0.0548 
\label{eqn:merr}
\end{equation}

The equation is valid for stars with $\log~g$ $<$ 3.5.
The RMS of the $\alpha$ abundances, shown as squares in the left panel of Figure \ref{fig:err},
are relatively low, of the order of 0.1 dex. Similarly to the metallicity, the 
$\alpha$ precision gets worse with decreasing metallicity due to the metallicity-$\alpha$ correlation found in the 
globular clusters. Near solar metallicities, the typical error spans between 0.02 and 0.07~dex, with no systematic
deviations as a function of metallicity or temperature. Because the RMS increases due to systematics in the 
metallicity-$\alpha$ correlation, we do not fit the RMS the of $\alpha$ abundances. We believe that the true random 
error of $\alpha$ is smaller than what is presented for the metal-poor stars.

\subsection{Future Improvements}

We plan several improvements in how we will obtain the physical parameters for APOGEE's next data release 
(DR12, planned for December 2014). This section provides an overview on those ongoing developments. 

The model atmospheres used in the current (DR10) version of ASPCAP are those published by \citet{castelli01} based
on the ATLAS9 code \citep{kurucz05}. We have recently produced an updated grid of model atmospheres
based on the most recent version of ATLAS9 \citep{meszaros01}, including an updated H$_2$ line list. 
The new grid of models considers, in addition to the usual variations in metal abundance, variations in 
$\alpha$-element enhancement and carbon abundance [C/M], which will lead to a spectral library with consistent
abundances between the synthesis and the model atmospheres. 

The [$\alpha$/M]$-$[M/H] correlation found in metal-poor and metal-rich clusters can be improved by separating iron
lines from OH lines and fitting them separately. This will lead to a more simple fit, where in the first step the 
metallicity 
is determined from iron lines, and in a second step $\alpha$ abundances can be derived 
from OH, and other $\alpha$-element lines.

We have clearly shown that our spectroscopic surface gravities differ systematically from those derived 
from isochrones or stellar oscillations. One or several of the many approximations involved in the calculation
of model spectra are likely 
behind these systematic errors. Departures from LTE, enhanced by
the low densities found in the atmospheres of giant stars, could be affecting some of the atomic populations. 
Molecular populations are quite sensitive to the thermal structure in high atmospheric layers,
and also to thermal inhomogeneities that must be present in the real stars, 
and these are completely neglected in our 1D model atmospheres, and the main
suspect for the observed systematic offsets. 

We have already started to look at how to bring ASPCAP gravities into better agreement with asteroseismic gravities. 
Grids of hydrodynamical model atmospheres are becoming 
available, and we plan to examine their impact on the H-band spectrum of giants in the near future 
\citep{chiavassa01, collet01, freytag01, ludwig01}. We have also noticed that the Brackett series 
lines in the spectra, which are fairly sensitive to pressure, are not always well matched by our model 
spectra, and their modeling may need to be revised.

While the overall size of the differences between our spectroscopic gravities and those from isochrones 
or pulsations are similar, we have identified some discrepancies when comparing ASPCAP surface gravities with these
two independent sources. We will examine closely these inconsistencies to identify 
the reasons behind them. Among the candidates we plan on exploring are the effects of the mixing length and
other convection  parameters as well as the helium content in the
construction of stellar  evolution models. We will also closely evaluate the applicability of the adopted scaling
relationship between the frequency of maximum power of oscillations ($\nu_{\rm max}$) and the surface gravity.

The ASPCAP analysis has so far been restricted to the main atmospheric parameters and the abundances of 
C, N and overall abundances of $\alpha$-elements and metals. We are further developing ASPCAP to handle 
the derivation of abundances for many more elements accessible from the H-band. Our preliminary studies
indicated that up to 15 different elements are well represented in this spectral window for 
late-type giant stars \citep{eis11}.

As part of these improvements, a manual analysis is being carried out for a number of the ASPCAP 
calibration clusters. These analyses will include derivations of the fundamental atmospheric 
parameters themselves, as well as measuring the additional 15 elements that are part of the APOGEE 
survey plan. These results will be used to further refine and calibrate the ASPCAP pipeline.

\section{Summary}

We have used data of 559 stars 
belonging to 20 Galactic globular and open clusters or with high-quality asteroseismic 
data to determine the accuracy 
and precision of atmospheric parameters published in DR10. The APOGEE spectra were run through the current 
version of the APOGEE Stellar Parameters and Stellar Abundances Pipeline (ASPCAP), 
and cluster members were selected carefully to compare effective temperature, metallicity, surface gravity, 
$\alpha$-elements, carbon and nitrogen abundances to literature values.

After carefully examining all six derived parameters we concluded the following:

Effective temperature: Effective temperatures agree well with spectroscopic temperatures from the literature 
within a mean offset of 8$\pm$161~K. For photometric temperatures, larger systematic differences can 
be seen. A correction function was provided to convert ASPCAP temperatures to 
the photometric scale of \citet{gonzalez01}, which is based on the IRFM method. 
The precision of ASPCAP temperatures was estimated using the RMS scatter of the clusters, and it is found to be
about 150$-$200~K in globular clusters, and 50$-$100~K in open clusters at solar metallicities. 

Metallicity: ASPCAP metallicities agree very well (to within 0.1~dex) with literature values for stars with 
$-0.5 <$ [M/H] $< +0.1$~dex. At both the metal-poor and metal-rich ends of the scale, systematic 
differences are apparent, amounting up to 0.2$-$0.3~dex. An empirical correction is provided using literature 
cluster averages to bring ASPCAP metallicities 
into agreement with the literature values. The metallicity scatter in each individual cluster is usually less 
than 0.15~dex, while the average scatter  in all 20 clusters is 0.09~dex. 

Gravity: ASPCAP gravities are larger by about 0.2$-$0.3~dex than both isochrones and seismic values 
for metallicity in the range $-0.5 <$ [M/H] $< +0.1$~dex. At lower
metallicities the difference is larger. An empirical correction is provided based on a combined 
data set including isochrone comparisons for metal-poor stars and \textit{Kepler} seismic values 
at solar metallicities. 

$\alpha$ abundances: [$\alpha$/M] abundances for stars within $-0.5 <$ [M/H] $< +0.1$~dex show no dependence 
on temperature, or metallicity. However, a clear correlation exists with [M/H] and T$_{\rm eff}$ outside 
this metallicity range, thus we advice to use $\alpha$ abundances with caution for stars with $-0.5 >$ [M/H] 
and $> +0.1$~dex. The typical precision in the above mentioned region is less than 0.07~dex. In addition, 
we are not confident on the derived [$\alpha$/M] results for $T_{\rm eff} lessim 4200$~K.

Carbon and nitrogen: The carbon and nitrogen abundances show significant 
systematic differences  (up to 1~dex) compared to literature values.
We currently discourage any use of those values in scientific applications. 

Further developments of ASPCAP will be available in the near future and aim to reduce the systematic effects seen 
in metallicity and gravity. The new data set with improved atmospheric parameters will be available in 
DR12, at the end of 2014. Continued improvements in the processing and analysis of APOGEE spectra will no doubt 
be one of the highest priorities for the APOGEE team in the upcoming years.

\acknowledgements{We thank Sara Lucatello, Joris De Ridder, Robert O'Connell, Leo Girardi, and Benoit 
Mosser for their detailed comments about the presented results. Saskia Hekker 
acknowledges financial support from the 
Netherlands Organisation for Scientific Research (NWO). Katia Cunha acknowledges support for this 
research from the National Science Foundation (AST-0907873). 
Verne V. Smith acknowledges partial support for this research from the National Science Foundation (AST-1109888).
Yvonne Elsworth and William J. Chaplin acknowledge support from STFC (The Science and 
Technology Facilities Council, UK). Sarbani Basu acknowledges NSF grant AST-1105930 and NASA ADAP grant  NNX13AE70G. 
This work has made use of BaSTI web tools.

The authors thankfully acknowledge the technical expertise and assistance provided by the Spanish Supercomputing 
Network (Red Espanola de Supercomputacion), as well as the computer resources used: the LaPalma Supercomputer, 
located at the Instituto de Astrofisica de Canarias. The authors also acknowledge the Texas Advanced Computing 
Center (TACC) at The University of Texas at Austin for providing H.P.C. resources that have con-
tributed to the research results reported within this paper (http://www.tacc.utexas.edu). 

Funding for SDSS-III has been provided by the Alfred P. Sloan Foundation, the Participating Institutions, the 
National Science Foundation, and the U.S. Department of Energy Office of Science. The SDSS-III web site is 
http://www.sdss3.org/.

SDSS-III is managed by the Astrophysical Research Consortium for the Participating Institutions of the SDSS-III 
Collaboration including the University of Arizona, the Brazilian Participation Group, Brookhaven National Laboratory, 
University of Cambridge, Carnegie Mellon University, University of Florida, the French Participation Group, the 
German Participation Group, Harvard University, the Instituto de Astrofisica de Canarias, the 
Michigan State/Notre Dame/JINA Participation Group, Johns Hopkins University, Lawrence Berkeley National Laboratory, 
Max Planck Institute for Astrophysics, New Mexico State University, New York University, Ohio State University, 
Pennsylvania State University, University of Portsmouth, Princeton University, the Spanish Participation Group, 
University of Tokyo, University of Utah, Vanderbilt University, University of Virginia, University of Washington, 
and Yale University. 

}

\end{document}